\documentclass[12pt,preprint]{aastex}

\newcommand{\degree}{\ensuremath{^\circ}}

\shorttitle{Beta Lyrae}
\shortauthors{Lomax et al.}

\begin{document}

\title{Geometrical Constraints on the Hot Spot in Beta Lyrae}


\author{Jamie R. Lomax} 
\affil{University of Denver, Department of Physics \& Astronomy}
\affil{2112 E. Wesley Ave, Denver, CO, 80208, USA}
\email{Jamie.Lomax@du.edu}

\author{Jennifer L. Hoffman} 
\affil{University of Denver, Department of Physics \& Astronomy}
\affil{2112 E. Wesley Ave, Denver, CO, 80208, USA}
\email{Jennifer.Hoffman@du.edu}

\author{Nicholas M. Elias II}
\affil{National Radio Astronomy Observatory}
\affil{Science Operations Center, PO Box O, 1003 Lopezville Road, Socorro, NM, 87801-0387, USA}
\email{nelias@nrao.edu}

\author{Fabienne A. Bastien\altaffilmark{1}}
\affil{Vanderbilt University, Department of Physics \& Astronomy}
\affil{6301 Stevenson Center, VU Station B \#351807, Nashville, TN 37235}
\email{fabienne.a.bastien@vanderbilt.edu}
\altaffiltext{1}{Fisk University, Department of Physics, 1000 17th Ave. N., Nashville, TN 37208}

\and

\author{Bruce D. Holenstein}
\affil{Gravic, Inc.}
\affil{301 Lindenwood Drive Suite 100, Malvern, PA 19355-1772 USA}
\email{BHolenstein@gravic.com}


\begin{abstract}
We present results from six years of recalibrated and new spectropolarimetric data taken with the University of Wisconsin's Half-Wave Spectropolarimeter (HPOL) and six years of new data taken with the photoelastic
modulating polarimeter (PEMP) at the Flower and Cook Observatory. Combining these data with polarimetric data from the literature allows us to characterize the intrinsic \textit{BVRI} polarized light curves. A repeatable discrepancy of 0.245 days (approximately 6 hours) between the secondary minima in the total light curve and the polarization curve in the \textit{V} band, with similar behavior in the other bands, may represent the first direct evidence for an accretion hot spot on the disk edge.
\end{abstract}





\section{INTRODUCTION}

Beta Lyrae A (also known as HD 174638, HR 7106 and ADS 11745A; hereafter ``$\beta$ Lyr") is a bright, well-studied semi-detached eclipsing binary star system. The primary star is a B6-B8 II, giant star (``loser") with a mass of 3 M$_{\sun}$ that is transferring matter to its main-sequence B0.5 V, 12.5 M$_{\sun}$, companion (``gainer") at about $10^{-5}$ M$_{\sun}$ yr$^{-1}$ via Roche lobe overflow \citep{HubenyPlavec,Harmanec1993}. This process has created a thick accretion disk that obscures the gainer \citep{Huang,Wilson,HubenyPlavec,Skulskii}. A bipolar flow or jet has also been detected in the system through interferometric and spectropolarimetric methods \citep[hereafter HNF]{Harmanec1996, Hoffman1998}. The system's mass ratio, $q$, has been placed between 4.2 and 6 with an inclination angle, \textit{i}, of 85$\degree$ \citep{Wilson}. Other studies have suggested $i=83 \degree$ with $q=5.6$ and $i=80 \degree$ with $q=4.28$ \citep{HubenyPlavec,Skulskii}. More recent determinations of the orbital inclination place its value at $i=86 \degree$ \citep{Linnell1998,Linnell2000}. These large mass ratios are evidence for mass reversal in the system's history. The disk's ability to obscure the gainer is due to the nearly edge-on inclination angle of the system. 

The system has a well-established orbital period of 12.9 days that increases at a rate of 19 s yr$^{-1}$ \citep{Harmanec1993}. Recent interferometric observations have produced the first images of the system, which show the loser and the disk as separate objects and confirm the orientation of the system axis, near $254\degree$, previously inferred from HNF's polarimetric analysis \citep{Harmanec1996, Zhao, Schmitt}. Understanding how mass moves between and around the stars and leaves the system is imperative to understanding the evolutionary future of $\beta$ Lyr. However, interferometric techniques have yet to resolve the mass stream or bipolar outflows. We have used spectropolarimetry to study the system and begin to unlock the evolutionary clues contained within the circumstellar material.

Light scattering from electrons in the highly ionized circumstellar material in $\beta$ Lyr produces a variable phase-dependent polarization. Since electron scattering preserves information about the orientation of the scattering region, analyzing polarimetric behavior as a function of wavelength allows us to determine from where in the system different spectral features have arisen. In this way, spectropolarimetric observations of $\beta$ Lyr can be used to infer the geometrical properties of the scattering material in the system. 

Optical polarimetry was used to study $\beta$ Lyr as early as 1934, but it was not known until 1963 that the system exhibited variable polarization \citep{Ohman,Shakhovskoi}. Appenzeller and Hiltner (1967; hereafter AH) were the first to publish interstellar polarization (ISP) corrected broadband \textit{UBV} polarization curves of $\beta$ Lyr. More recently, HNF published ISP-corrected polarized light curves in the \textit{V} band and H$\alpha$ and \ion{He}{1} $\lambda 5876$ emission lines using a subset of the data we present here. HNF used the position angles of the polarized UV continuum and the hydrogen Balmer emission lines to confirm that a bipolar outflow exists in the $\beta$ Lyr system after their discovery by Harmanec et al. (1996). HNF also interpreted the average position angle of the visible polarized light ($164\degree$) to be the physical axis of the binary system, an interpretation which was borne out by the interferometric images presented by Zhao et al (2008) and Schmitt et al. (2009).

In this paper we present new \textit{BVRI} and \ion{He}{1} $\lambda$5876 polarization curves and polarized light curves of $\beta$ Lyr. The details of our spectropolarimetric observations and our interstellar polarization corrections are in Section 2. Section 3 presents and displays our observational results. We analyze our findings in Section 4 and summarize conclusions in Section 5.

\section{OBSERVATIONS}

This study compiles data from three distinct data sets. The first consists of 69 optical spectropolarimetric observations of $\beta$ Lyr taken over 6 years with the University of Wisconsin's Half-Wave Spectropolarimeter (HPOL) at the 0.9 m telescope at Pine Bluff Observatory (PBO); the second data set comprises 6 years of broadband optical polarimetric data obtained at the Flower and Cook Observatory; and the third is 3 years of archived broadband optical polarimetric data from AH taken with the 24 inch rotatable telescope at the Yerkes Observatory. To calculate the phase for each observation, we used the ephemeris 
$$T_{\mathrm{pri}}=\mathrm{HJD}\: 2,408,247.966 + 12.91378E + 3.87196 \times 10^{-6}E^{2}$$
\noindent where \textit{E} is the total number of orbits since the primary eclipse that occurred at HJD 2,408,247.966 \citep{Harmanec1993}. This is the same ephemeris used by HNF; it does not significantly differ from the more recent ephemeris presented by Ak et al. (2007). 

\subsection{HPOL Data Set}

The first 14 HPOL observations, obtained between 1992 September and 1994 November, used a dual Reticon array detector with a wavelength range of 3200-7600 \AA \ and a resolution of 15 \AA \ (see Wolff, Nordsieck, \& Nook 1996 for further instrument information). The remaining 55 observations, taken between 1995 March and 1998 September, used a CCD-based system. This extended the wavelength range, 3200-10500 \AA, and increased the resolution to 7.5 \AA \ below 6000 \AA \ and 10 \AA \ above \citep{NordsieckHarris}. The first 29 observations were previously published in HNF; they have undergone recalibration for use in this study. 

Table \ref{betLyrObs} lists the orbital phases along with civil and heliocentric Julian dates for the midpoint of each HPOL observation. Each $\beta$ Lyr observation covers the full spectral range, with the exception of the four nights indicated in Table \ref{betLyrObs}. Two observations, 1995 May 27 and 1995 August 14, used only the red grating (6,000-10500 \AA) of the CCD system while the other two, 1997 May 17 and 1997 May 26, used only the blue grating (3200-6000 \AA). Each individual observation typically lasted between 45 minutes and an hour (approximately 0.03 to 0.04 days) when both gratings were used. 

We used 11 HPOL observations of $\beta$ Lyr B taken between 1995 May and 1999 November to obtain an ISP estimate. Beta Lyr B is a member of the same association as $\beta$ Lyr A and is located 45" away \citep{Abt}. All of these observations were made using HPOL's CCD-based system. Table \ref{betLyrObs} also lists the civil and heliocentric Julian dates that correspond to the midpoints of each observation of $\beta$ Lyr B and indicates which grating(s) were used during the observations. The first two observations, 1995 May 21 and 1996 July 3, were previously published in HNF and have undergone recalibration for this study. We reduced all of the HPOL observations using the REDUCE software package (described by Wolff et al. 1996). 

We estimated the ISP by fitting a modified Serkowski law curve to the error-weighted mean of the 11 observations of $\beta$ Lyr B \citep{Serkowski,Wilking}. The parameters for our ISP estimate are $P_{\mathrm{max}}=0.422\% $ $\pm$ $0.005\%$, $\lambda_{\mathrm{max}}=4149$ \AA \space$\pm$ $80$ \AA, $K=0.699$ and $\mathrm{PA}=151.16\degree$ $\pm$ $0.36\degree$. We subtracted this ISP estimate from the HPOL $\beta$ Lyr data. This new ISP correction has significantly improved uncertainties over previous estimates; it is consistent with the HNF estimate ($P_{\mathrm{max}}=0.419\% \pm 0.013\%$, $\lambda_{\mathrm{max}}=4605$ \AA \space$\pm$ $260$ \AA, and $\mathrm{P.A.}=151.0\degree$ $\pm$ $0.9\degree$), which was determined using only the 1995 May 21 HPOL observation. AH found a similar estimate, $P_{max}=0.42\%$ \space$\pm$ $0.04\%$ and $PA=153.2\degree$ $\pm$ $3\degree$ by taking the weighted mean of the observed polarization of the associated stars $\beta$ Lyr B, E, and F.

\subsection{FCO Data Set}

Our second data set is made up of 19 \textit{B} band, 88 \textit{V} band and 17 \textit{R} band observations obtained at the Flower and Cook Observatory between 1987 and 1992 with the PEMP instrument \citep{Bruce,Elias}. The length of each observation was between 20 and 25 minutes (approximately 0.01 days). The phases along with the civil and heliocentric Julian dates for each observation are listed in Table \ref{betLyrObsFCO}. We have no observations of $\beta$ Lyr B taken with the same instrument as this data set. Therefore, we used the Serkowski fit to the HPOL $\beta$ Lyr B observations (see Section 2.1) to calculate the ISP contributions at the central wavelengths of the \textit{BVR} bands and subtracted these estimated values from the observations in this data set. We list these data and ISP subtracted data in Tables \ref{BFCOData} through \ref{RFCOData}.

\subsection{AH Data Set}

We also used archival \textit{BV} polarization data taken between 1964 and 1966, originally published in AH. This data set consists of 37 \textit{B} band and 127 \textit{V} band observations, the details of which can be found in AH (and references therein). The DC two-channel polarimeter was rotated $30\degree$ between 12 separate 20 second exposures of $\beta$ Lyr and the sky \citep{Appenzeller}. Therefore, the total integration time for both $\beta$ Lyr and the sky was 4 minutes (approximately 0.003 days). We converted these data from polarization magnitudes to percent polarization and converted their Julian dates to heliocentric Julian dates for use in this study.  For consistency between the HPOL and AH data sets, we did not use AH's published ISP corrected data because it included two stars ($\beta$ Lyr E and F) that are not included in the HPOL ISP estimate. Instead, we subtracted only the AH \textit{BV} $\beta$ Lyr B observations from their non-ISP corrected $\beta$ Lyr data in each respective band.

\section{RESULTS}

\subsection{Broadband Polarimetry} 
 
\begin{figure}
\figurenum{1}
\epsscale{1.0}
\plotone{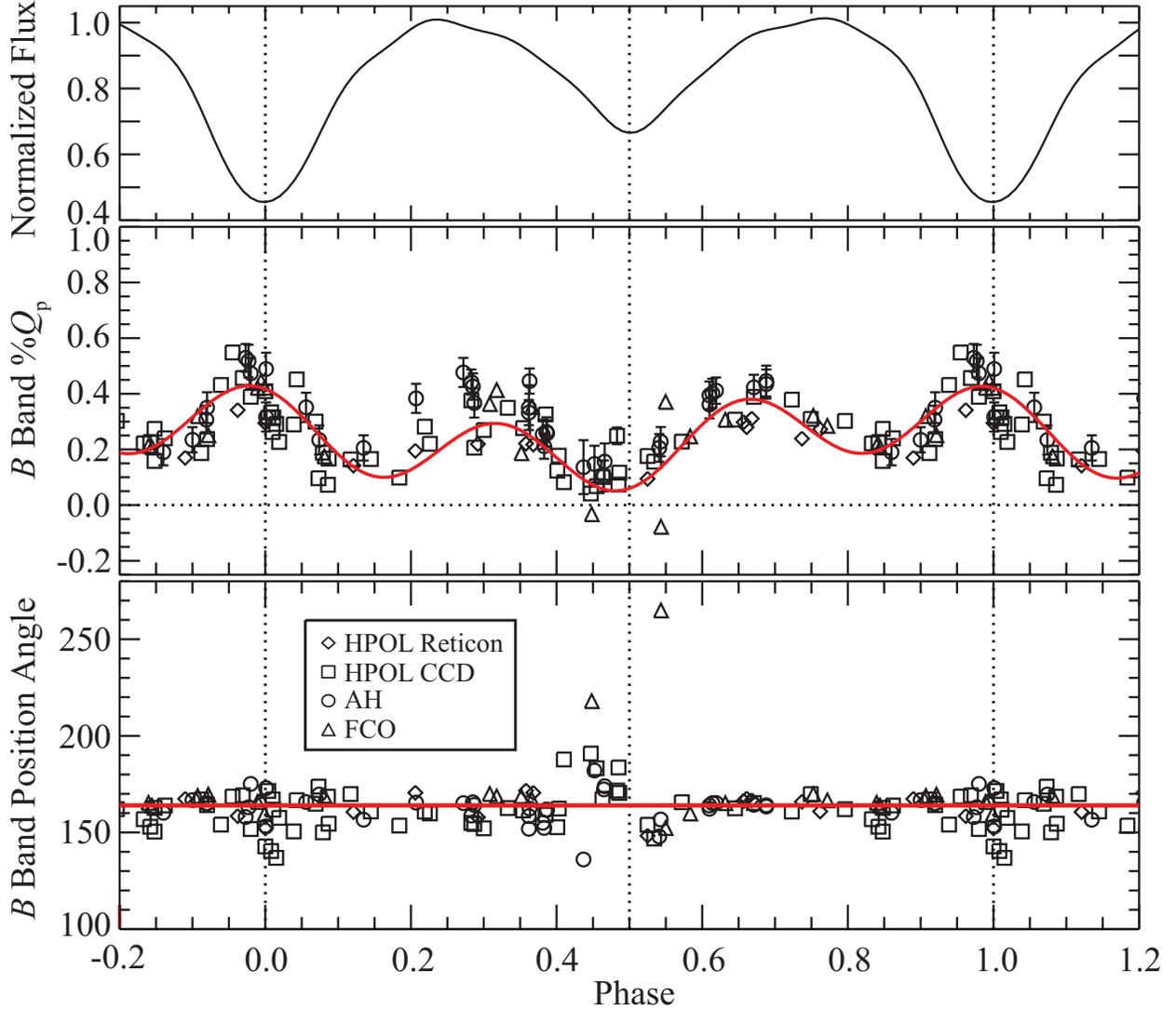}
\label{Bband}
\caption{Data points represent the \textit{B} band polarimetric observations from HPOL Reticon (diamonds), HPOL CCD (squares), AH (circles) and FCO (triangles). \textit{From top:} Normalized \textit{V} band Fourier fit light curve \citep{Harmanec1996}, projected polarization (see Section 3.1), and position angle (degrees) versus phase. Error bars are shown for uncertainties larger than 0.025 in $\%\textit{Q}_p$ and $5.0\degree$ in position angle. The HPOL error bars shown represent the larger of the intrinsic and systematic uncertainties. All data have been wrapped so that more than one complete period is shown. The solid line in the middle panel represents our Fourier fit to the $\%\textit{Q}_p$ data (see Section 3.1). The dotted line represents zero projected polarization. The solid line in the bottom panel represents the average position angle of $164\degree$.}
\end{figure}

\begin{figure}
\figurenum{2}
\epsscale{1.0}
\plotone{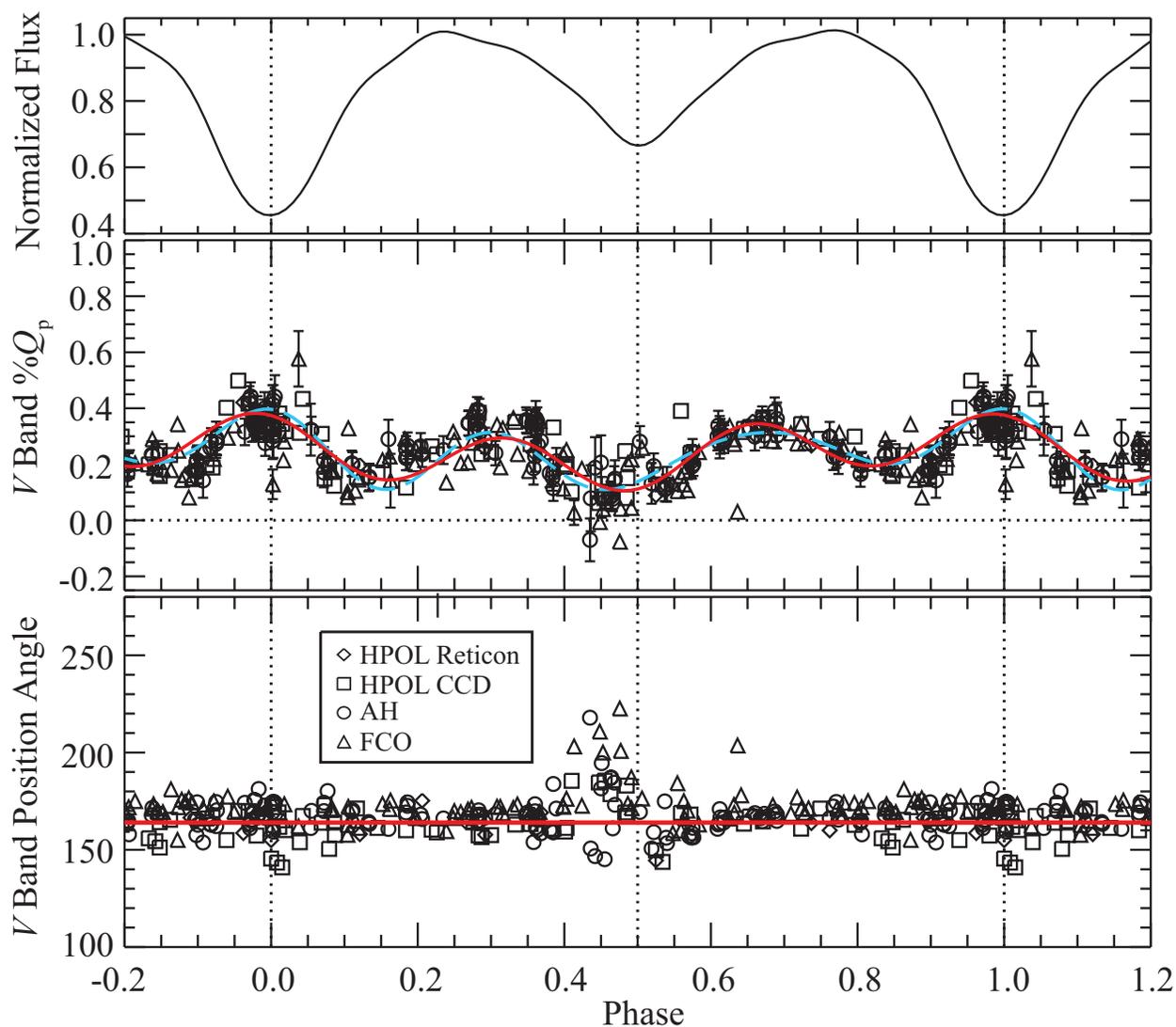}
\label{Vband}
\caption{Same as Figure 1, but for \textit{V} band polarimetry. The dashed blue curve (in the online version, otherwise dashed light grey) represents the three-term Fourier fit and the red curve (in the online version, otherwise solid) represents the two-term Fourier fit. Note the two fits differ at primary and secondary eclipse, the quadrature phases, and near phases 0.15 and 0.9.}
\end{figure}

\begin{figure}
\figurenum{3}
\epsscale{1.0}
\plotone{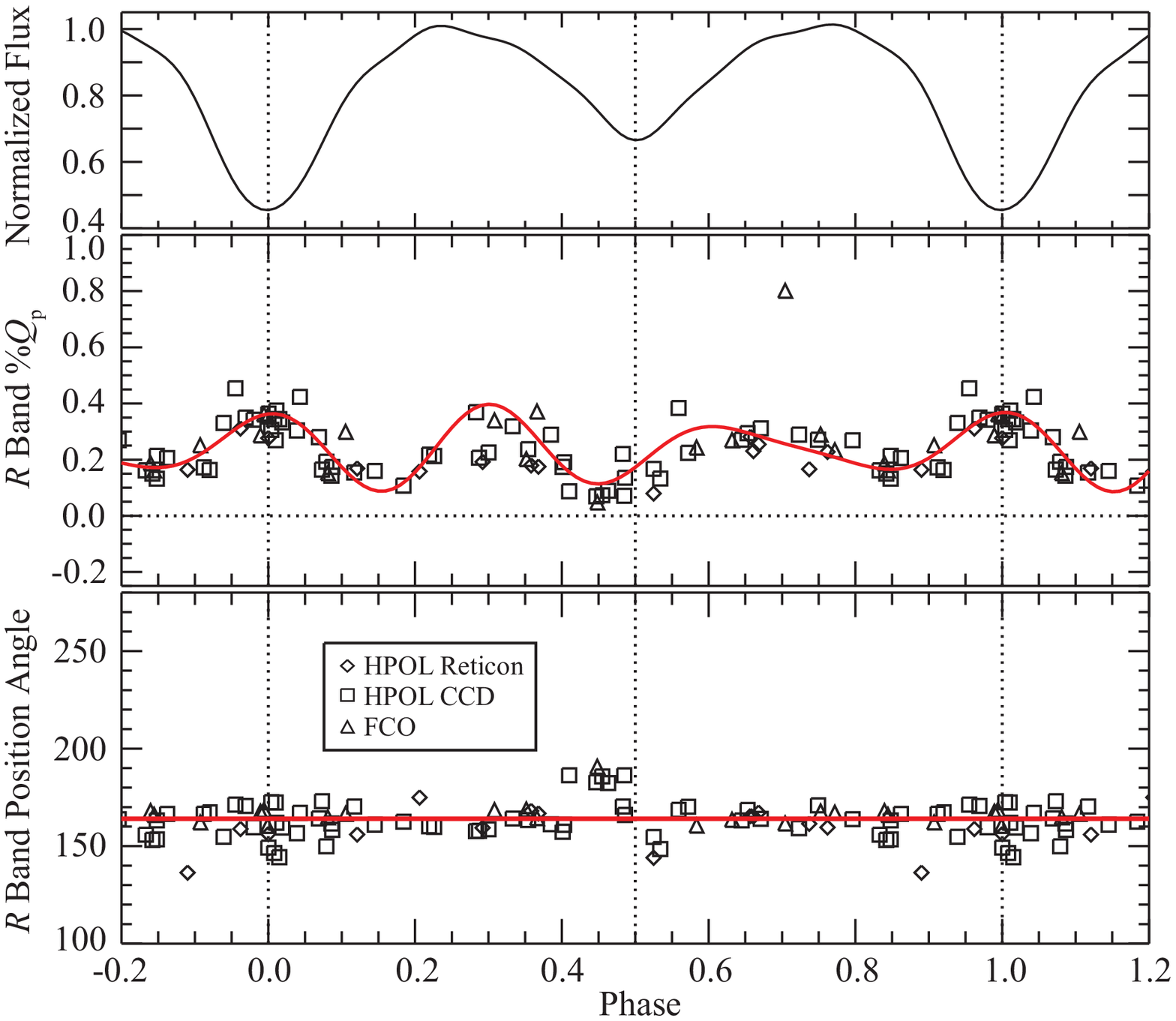}
\label{Rband}
\caption{Same as Figure 1, but for \textit{R} band polarimetry.}
\end{figure}

\begin{figure}
\figurenum{4}
\epsscale{1.0}
\plotone{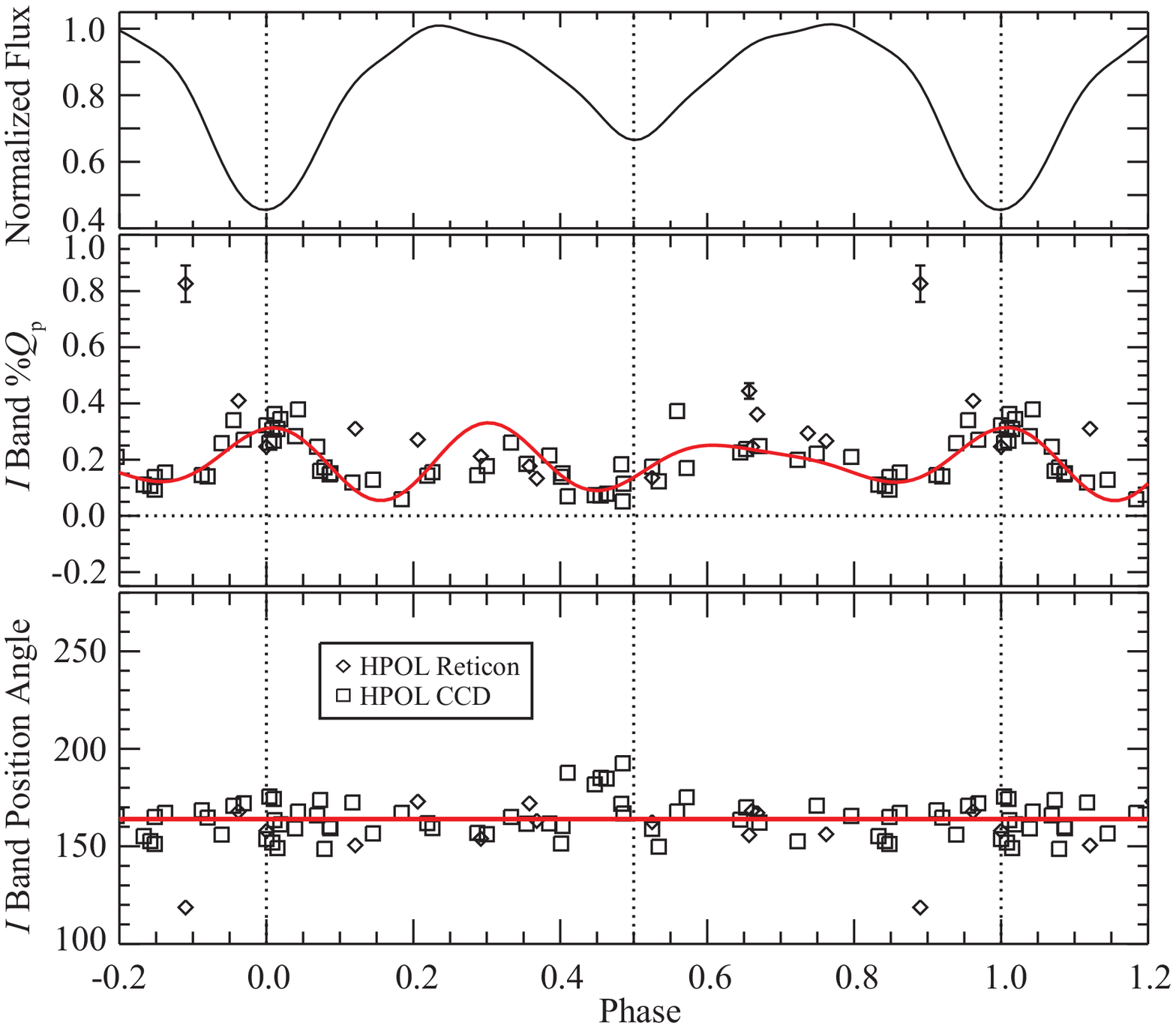}
\label{Iband}
\caption{Same as Figure 1, but for \textit{I} band polarimetry.}
\end{figure}

To examine the behavior of the continuum polarization with orbital phase, we applied synthetic \textit{BVRI} Johnson-Cousins band filters (described by Bessell 1990) to the ISP-corrected HPOL data. The filter routine produces broadband values and associated internal errors for each observation; however, we must still take into account systematic variations in the instrumental polarization between nights. Systematic errors for HPOL at PBO were evaluated by periodically analyzing observations of unpolarized standard stars. Tables \ref{BHPOLData} through \ref{IHPOLData} list broadband polarization values and internal errors determined by the filter routine along with the systematic errors. In the case of the Reticon data, the systematic errors are less well determined; based on our previous experience with these data, we have estimated the uncertainties in the Stokes parameters for the Reticon data to be $0.02\%$ in all bands. Figures \ref{Bband} through \ref{Iband} display these data graphically, using the larger of the internal and systematic errors for each observation, along with the FCO and AH data.

The position angles for the \textit{BVRI} bands remain relatively constant with orbital phase except near secondary eclipse, where the position angle values appear to rotate away from the mean value. The bottom panels in Figures \ref{Bband} through \ref{Iband} show this behavior in each band. In the standard picture of the system, the polarization is produced by scattering from the accretion disk edge; the average position angle at these wavelengths should therefore provide us with an estimate of the orientation of the axis of the disk and thus of the system as a whole. To calculate the average position angle in each band, we excluded the secondary eclipse points (between phases 0.425 and 0.575) because they do not follow the near-constant trend displayed at other phases and performed a linear, error-weighted, least-squares fit to the remaining data in \textit{Q-U} space. We then used the slopes of the fitted lines to determine the position angles listed in Table \ref{betLyrPA}. We also calculated a position angle for Balmer jump index (the vector difference between the polarization above and below the Balmer jump). As discussed in HNF, the broadband polarization in $\beta$ Lyr undergoes a $90\degree$ position angle rotation across the Balmer Jump. Thus, this vector difference defines the system axis in \textit{Q-U} space. Because the Balmer jump index is independent of the ISP, it provides us with an independent estimate of the orientation of the system axis. 

We find the weighted mean position angle of the HPOL Reticon and CCD bands and the Balmer jump index to be $ 164.6\degree \pm 0.22\degree$. For each band, the CCD and Reticon position angles do not agree within uncertainties, with the Reticon data yielding larger position angles in all bands. This result is due to poor sampling in \textit{Q-U} space; the Reticon data only consist of 14 data points in each band while there are 55 observations in the CCD \textit{V} and \textit{R} bands and 53 observations in the CCD \textit{B} and \textit{I} bands. This skews the linear fit since the full range of possible observable \textit{Q-U} values is not well covered by the Reticon data. However, the larger systematic uncertainties we adopt for the Reticon data result in these points carrying a lower weight in the fit; thus, we are confident that our weighted mean is a fair representation of the true system axis. HNF found the mean \textit{V} band position angle to be $163.8\degree \pm 0.15\degree$; while our estimate is not formally consistent with HNF's, it shows the broad band polarization behavior of the system is the same in all optical bands. Our polarization position angle implies a system position angle, defined by the position angle of the disk axis, of $253.8\degree \pm 0.15\degree$ on the sky. As expected for polarization by electron scattering, the position angle of the polarized light is perpendicular to the position angles describing the system orientation given by Zhao et al. (2008) and Schmitt et al. (2009). Zhao et al. (2008) estimated the position angle of the system's ascending node as $253.22\degree \pm 1.97\degree$ and $251.87\degree \pm 1.83\degree$ using two different image reconstruction techniques on their interferometric data and $254.39\degree \pm 0.83\degree$ using a model of the system, while Schmitt et al. (2009) estimate $249.0\degree \pm 4.0\degree$.    

We rotated all of the HPOL, AH and FCO data to the average position angle of $164\degree$. This orients our data with respect to the intrinsic polarization axis of the system. After this rotation, $\%\textit{U}$ averages to zero in each band and the polarization varies significantly only in the $\%\textit{Q}$ direction. In the rest of this paper we present the projected Stokes parameter $\% \textit{Q}_p$ resulting from this rotation. The use of this quantity is beneficial because it can be positive or negative, whereas \%\textit{P} is always positive. Data points that have a position angle near $164\degree$ will have a positive $\% \textit{Q}_p$ value while points with a position angle perpendicular to this (near $74\degree$) will have a negative $\% \textit{Q}_p$. Hereafter we display only $\% \textit{Q}_p$ because $\% \textit{U}_p$ values scatter around zero. Because the rotation is a simple trigonometric calculation, we present in Tables \ref{BFCOData} through \ref{IHPOLData} the unrotated $\% \textit{Q}$ and ISP subtracted FCO and HPOL data only. 

The middle panels in Figures \ref{Bband} through \ref{Iband} show the $\% \textit{Q}_p$ curves for the \textit{BVRI} bands after rotation. We used the program PERIOD04 to perform a Fourier fit to the data for each band \citep{Period04}. The PERIOD04 fitting formula is $y= Z + \sum_{i=1}^n A_i \sin(2\pi (\Omega_i t + \phi_i))$ where \textit{n} is the number of sine terms in the fit, \textit{Z} is the zero point, \textit{A} is the amplitude, $\Omega$ is the frequency, and $\phi$ is the phase. The results of the fits are displayed as solid curves in each figure and their parameters are given in Table \ref{FitParam}. The \textit{V} band, and to a lesser extent the \textit{B} band, Fourier fits deviate from the data at phase 0.9 (see Figure \ref{Vband}). This discrepancy disappears in the \textit{V} band if we include three frequency terms in the Fourier fit. However, we are not as confident in the third frequency as we are in the first two because the PERIOD04 fitting program produces a reasonable third term for the \textit{V} band only. Therefore, in Table \ref{FitParam} we report parameters for only the first two terms, but we display both the two-term and three-term fits in Figure \ref{Vband}. These fits provide the first quantitative representations of the polarization variations in the $\beta$ Lyr system.

The data in the \textit{BVRI} bands are almost always positive, indicating their position angles stay near $164\degree$ throughout the orbital period. Each of the bands displays an increase in $\% \textit{Q}_p$ at primary eclipse and two other increases near the quadrature phases (0.25 and 0.75). The height difference between the polarization bumps at the quadrature phases noted by HNF disappears now that more data are added, but we note that the \textit{B} and \textit{V} phase 0.25 bump have a higher dispersion around the average $\% \textit{Q}_p$ value than does the 0.75 phase bump. The \textit{R} and \textit{I} band show the opposite behavior; future observations will be able to tell us whether this is due to their poor phase coverage or whether it indicates that the \textit{R} and \textit{I} bands are probing a different region of the disk than the \textit{B} and \textit{V} bands. We calculated the variance of the two quadrature bumps between phases 0.25 and 0.35, and 0.65 and 0.75 to formally show this. The first quadrature bump has \textit{BVRI} variances of 0.081 $\pm$ 0.006, 0.110 $\pm$ 0.005, 0.040 $\pm$ 0.002, and 0.016 $\pm$ 0.002 respectively, while the second quadrature bump has variances of 0.048 $\pm$ 0.005, 0.027 $\pm$ 0.002, 0.281 $\pm$ 0.005, and 0.069 $\pm$ 0.004. 

The \textit{R} and \textit{I} bands also produce a lower polarization signal than the \textit{B} and \textit{V} bands. This change in polarization behavior with wavelength could indicate that scattering mechanisms other than electron scattering are present in the system. However, more observations in the \textit{R} and \textit{I} bands are needed to rule out the possibility that the low signal is due to a lack of phase coverage in these bands.

Figures \ref{Bband} through \ref{Iband} also show that in each band, the polarization curve has a fitted minimum in polarization that occurs just prior to the secondary eclipse in total light. This minimum is accompanied by a rotation in the position angle of the polarized light away from its average value. Table \ref{HSSize} gives the phases of the $\%\textit{Q}_P$ Fourier fit minimum in all bands. The offset between secondary eclipse in polarized light and total light is a new result, seen here for the first time due to the improved phase coverage in these data. We discuss the implications of this phenomenon in Section 4.

At primary eclipse, there are hints of similar behavior: the polarization maxima in the \textit{B} and \textit{V} bands occur slightly before phase 0.0, and all bands show a deviation from the mean position angle at and just after phase 0.0. However, we consider the primary eclipse features to be less significant than the ones at secondary eclipse for the following reasons. The three data points showing noticeable deviations from the mean position angle all occurred on the same night, 1997 August 25, which suggests that this effect may be due to a non-periodic process intrinsic to the $\beta$ Lyr system, or to a change in observing conditions that affected that night's data. If the same structure is responsible for the phenomena at both eclipses, we expect that a similar position angle scatter should exist in observations from the same orbit of the system whose phases are between primary and secondary eclipse, when such a structure should be most visible to the observer. However, neither the data from 1997 August 26, nor 1997 August 30 show such behavior. Additionally, the polarization maximum in the \textit{R} band occurs at phase 0.0 within the uncertainties, and in the \textit{I} band the maximum occurs just after phase 0.0. If we include the third term in the \textit{V} band Fourier fit, the primary polarization maximum occurs at phase 0.0 within uncertainties. Because the eclipse behavior is not consistent between bands and the associated position angle scatter appears to have occurred during only one orbit of the system, we consider it unlikely that these are due to a stable physical structure within the system (see Section 4 for further discussion of the effects seen at primary eclipse). 

\begin{figure}
\figurenum{5}
\epsscale{1.0}
\plotone{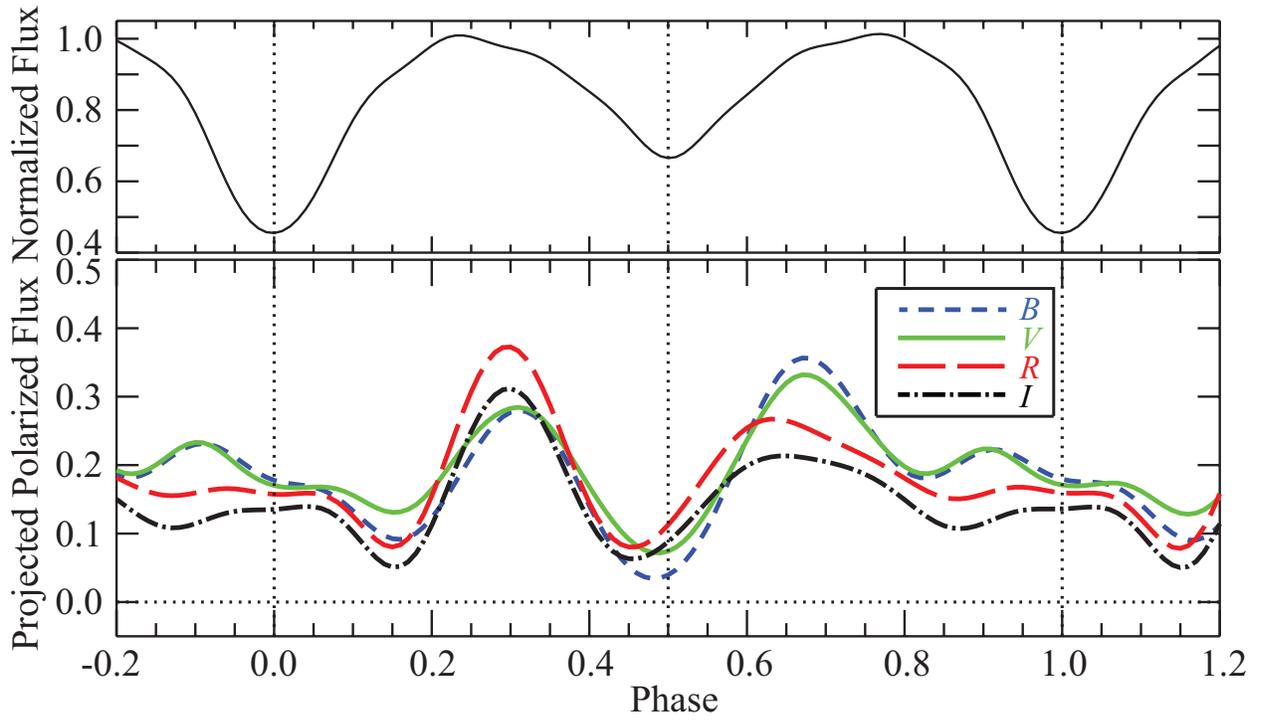}
\label{allbands}
\caption{\textit{From top:} Normalized \textit{V} band Fourier fit light curve and projected polarized flux curves for the \textit{BVRI} bands (see Section 3.1). The projected polarized flux curves are formed by multiplying each band's Fourier fit polarization curve by its normalized to maximum light Fourier fit light curve. }
\end{figure}

Figure \ref{allbands} displays the projected polarized \textit{BVRI} flux light curves for the $\beta$ Lyr system. To create these polarized light curves, we multiplied the fitted polarization curves shown in Figures \ref{Bband} through \ref{Iband} by their respective Fourier fit light curves \citep{Harmanec1996} normalized to maximum light. In all bands, the polarized flux remains nearly constant across primary eclipse due to a decrease in total light and an increase in $\%\textit{Q}_P$. The secondary eclipse offset seen in Figures \ref{Bband} through \ref{Iband} persists in Figure \ref{allbands}, while quadrature phases display local maxima. The \textit{B} and \textit{V} band appear to be the most similar; they overlap for most phases, while outside of secondary eclipse and the first quadrature phase the \textit{R} and \textit{I} bands produce the lowest net projected polarized flux. We do not consider the apparent height differences between bands at the quadrature phases to be significant due to the scatter in the observational points and the lack of coverage in the \textit{R} and \textit{I} bands.

\subsection{Line Polarimetry} 

We also took advantage of the spectropolarimetric nature of the data by studying the polarization behavior of the strongest optical emission lines in $\beta$ Lyr's spectrum: H$\alpha$, H$\beta$, \ion{He}{1} $\lambda$5876, \ion{He}{1} $\lambda$6678, and \ion{He}{1} $\lambda$7065. HNF hypothesized that the H$\alpha$, H$\beta$, \ion{He}{1} $\lambda$5876 and \ion{He}{1} $\lambda$7065 lines, which show a negative projected polarization, scatter in the bipolar outflow, while the \ion{He}{1} $\lambda$6678 scatters on the edge of the disk. However, HNF did not have enough data to construct a full polarization phase curve for the lines. Our expanded data set allows us to do this. However, we present only our \ion{He}{1} $\lambda$5876 results in graphical form because our uncertainties are relatively large due to signal-to-noise limitations. Rather than present the H$\alpha$, H$\beta$, \ion{He}{1} $\lambda$6678, and \ion{He}{1} $\lambda$7065 data, we describe their general behavior below. Future observations will allow us to use these data to draw quantitative conclusions about the scattering regions that give rise to the polarization in these lines. 

In order to calculate the polarization for each emission line, we used the flux equivalent width method described by HNF, using the same line and continuum regions as far as possible. We corrected the H$\alpha$ and H$\beta$ lines for underlying unpolarized absorption components (arising from the loser) in the same manner as HNF, using their preferred absorption equivalent widths of $8\pm 2$ \AA \space for H$\alpha$ and $6\pm 1$ \AA \space for H$\beta$. This has the following effect on the data. The continuum is positively polarized while the lines are negatively polarized. If we do not correct for unpolarized absorption, we remove too much continuum, and thus our resulting line polarization is too negative. With the absorption correction, the continuum contribution is smaller and the magnitude of the polarization is also smaller, resulting in a less negative $\%\textit{Q}_p$.

We do not present the HPOL Reticon line polarization values due to their large uncertainties. Figure \ref{HeI5876} shows $\%\textit{Q}_p$ and position angle curves for the \ion{He}{1} $\lambda$5876 line. It has a negative $\%\textit{Q}_p$; thus its position angle is perpendicular to the intrinsic axis of the system. In addition, the polarization for the line approaches zero at both primary and secondary eclipses. For this to happen, the scattering region for this line must lie near enough to the orbital plane of the system to be occulted both by the loser and by the disk. HNF previously suggested that this line scatters in the bipolar outflows because its average position angles lie near $74\degree$, corresponding to negative values of $\%\textit{Q}_P$. The results from our extended data set support this interpretation and further suggest that the \ion{He}{1} $\lambda$5876 scattering region within the outflows must lie between the loser and the disk and have a vertical extent comparable to the height of the disk.

The H$\alpha$, H$\beta$, and \ion{He}{1} $\lambda$7065 lines also all display a negative $\%\textit{Q}_p$ and are likely scattered in the same region as the \ion{He}{1} $\lambda$5876 line. Their average position angles are listed in Table \ref{betLyrPA}.  

The \ion{He}{1} $\lambda$6678 data show a polarization behavior different from that of the other lines. The data are generally positively polarized and their average position angle, $138.2\degree$, agrees more closely with the intrinsic axis of the system than do those of the other emission lines (see Table \ref{betLyrPA}).

\subsection{Period Analysis}

Besides the primary orbital period of $\beta$ Lyr (12.9 days), analysis of light curves has revealed several longer periodicities. A 340-day period was detected by Peel (1997), while both Van Hamme et al. (1995) and Harmanec et al (1996) detected a 282-day period. Wilson and Van Hamme (1999) searched polarimetry from AH, HNF, \cite{Serkowski}, and \cite{Shulov} for periodicities but did not detect anything significant. 

We performed a Lomb-Scargle power spectrum analysis to search for any periodic behavior in our polarimetric broadband (HPOL, AH, and FCO) and line data (H$\alpha$ and \ion{He}{1} $\lambda$ 5876) not associated with the 12.9 day orbital period of the system. While none of the previously detected longer periods were found, our analysis indicates the presence of periods of approximately 4.3 days in both the \textit{V} and \textit{B} bands with a False Alarm Probability (FAP; Horne \& Baliunas 1986) of $10^{-6}$ when using all three data sets. None of the other bands or the line polarization data appear to contain periods other than the orbital period. We also searched for periods within the HPOL, AH and FCO data sets individually to look for any transient periodic variations. We find that the 4.3 day period also appears in the \textit{V} and \textit{B} AH data set with a FAP of $10^{-6}$, but it does not appear in the other two data sets. This period is exactly one third of the 12.9-day orbital period of the $\beta$ Lyr system and results from the combination of two separate effects: the increase in polarization at the quadrature phases due to light scattering off the disk edge and the increase in polarization at primary eclipse due to the occultation of unpolarized light by the loser \citep[hereafter HWN]{Hoffman2003}. These two effects cause the $\%\textit{Q}_p$ curves to form a complete cosine curve between phases 0.0 and 0.3, a second cosine curve between phases 0.3 and 0.6 and a third cosine between 0.6 and 1.0 (see Figures 1 through 4). Therefore, this period does not provide new information about the $\beta$ Lyr system. 

We also performed a much simpler analysis to search for signatures of the 282-day period, which has been ascribed to variability of the conditions of the circumstellar and circumbinary gas \citep{Wilson, Ak}. Time plots of $\%\textit{Q}_p$, $\%\textit{U}_p$, position angle, and percent polarization for each of the \textit{BVRI} bands and the \ion{He}{1} $\lambda$5876 line did not reveal the 282-day period and are therefore not shown. However, the time coverage of our observations is very uneven and could have prevented us from detecting variations on this time scale. The observations in our data set were taken between 1964 and 1998, but we have no polarimetric observations taken during 1990, 1991, or between 1966 and 1987. The number of observations performed per year only adds further complications; several years have very few observations. We also note that the only the HPOL data set has observations of every band and the \ion{He}{1} $\lambda$5876 line. For these reasons, we also performed an analysis similar to Harmanec et al. (1996) on the $\%\textit{Q}_p$ data for each band and the \ion{He}{1} $\lambda$5876 line. Plotting the data on the 282.37 day period mitigates the lack of time coverage by ``folding" observations onto one cycle.

We plotted the data from selected orbital period phase bins versus their phase on the 282.37 day period (see Harmanec et al. 1996 for a similar analysis of a large amount of \textit{V} band total-light photometry). We chose the following five orbital phase bins for several reasons: the 0.0-1.0 bin allows us to use all of the data, the 0.6-0.15 bin is where Harmanec et al. (1996) detected the 282.37 day period the most strongly, the 0.25-0.35 and 0.65-0.75 bins allow us to determine whether the polarization of the quadrature phases changes on a 282.37 day time scale, and the 0.425-0.575 bin allows us to determine whether the secondary eclipse has a 282.37 day periodic behavior. The 0.25-0.35, 0.425-0.575, and 0.65-0.75 orbital period bins are too tight to leave a useful number of observations in the \textit{R} and \textit{I} bands and the \ion{He}{1} $\lambda$5876 line. While the \textit{B} and \textit{V} bands are slightly less affected by the size of the 0.25-0.35, 0.425-0.575, and 0.65-0.75 orbital phase bins, their behavior does not indicate, by eye or by using PERIOD04, the presence of a variation on a 282.37 day period. The size of the 0.6-0.15 and 0.0-1.0 bins present the best chance of detecting this period because the amount of data is not severely reduced. However, neither bin reveals the presence of the 282.37 day period. We also subtracted the Fourier fits from the data in the \textit{BVRI} bands (see Section 3.1) and searched the residuals for the 282.37 day period with the same bins used for the original $\%\textit{Q}_p$ data. The analysis on the residuals produced similar results; we do not find evidence of a 282.37 day period in either the $\%\textit{Q}_p$ data or the Fourier fit subtracted residuals. However, we note that the Fourier fits deviate from the data at some phases (see Section 3.1). Plots from this analysis resemble scatter plots and are therefore not shown.

\section{DISCUSSION}

In interpreting their polarized flux curves, HNF proposed two different possibilities for the origin and scattering location of the visible light. In their ``disk-disk" case, this light arises from within the disk and scatters from the disk edge; in the ``loser-lobe" case, the \textit{V}-band continuum light arises from the loser and scatters from material between the loser and the disk. In this analysis, HNF implicitly assumed that all features of the visible polarized flux curve are due to the same origin and scatterer. However, modeling work by HWN and subsequent modeling results (Hoffman et al., in prep.) have shown that the scattered light can originate both from the loser and from the disk in differing proportions over the binary cycle. These newer results suggest the following interpretations of our \textit{BVRI} polarization curves. The net increase in $\% \textit{Q}_p$ at primary eclipse (Figures \ref{Bband}-\ref{Iband}) is the result of the unpolarized light from the primary star being blocked by the disk material at phase 0.0. HNF interpreted the increase in $\% \textit{Q}_p$ at the quadrature phases as arising in one of two ways: light originating from within the disk and scattering from the disk edge, or light originating from the loser and scattering from material between the stars. We propose, based on recent modeling work by Hoffman, that these ``quadrature bumps" form simply by loser light scattering from the disk edge. The minimum at secondary eclipse occurs because the unpolarized primary star blocks light scattered in the secondary component.

Near secondary eclipse in all four broadband $\% \textit{Q}_p$ curves, the minimum in polarization precedes the minimum in total light; the phases for the polarization minimum in the \textit{BVRI} bands are listed in Table \ref{HSSize}. In each band, this minimum corresponds to a rotation in position angle away from the average value; the phase ranges for this rotation are also listed in Table \ref{HSSize}. In the basic star-star-disk model for the system, there is no mechanism to produce this disparity. If the loser is an unpolarized source, as indicated by the absence of a primary eclipse in the polarized flux curves (Figure 5; HNF), then the polarization minimum produced by its transit across the disk should be centered at flux minimum (phase 0.5; HWN). Thus, to explain this offset, we need to invoke another system component. Since $\beta$ Lyr is a mass transfer system, it most likely contains a mass stream connecting the loser and the disk as well as a `hot spot' where the mass stream from the loser interacts with the disk edge \citep[see also the geometries proposed by HNF]{Lubow}. Some studies (for example, Bisikalo et al. 2000 and references therein) suggest that the manner in which the mass stream approaches the disk prevents a hot spot from forming. Instead, a portion of the stream makes a full revolution around the disk and then interacts with the original stream. The process of this interaction allows the material that has made a full revolution around the secondary star to become part of mass stream again; Bisikalo et al. (2000) do not consider it to be part of the disk. Since this material's position angle is the same as the disk in the system, polarimetry cannot distinguish between the two possibilities. Therefore, we use the term `hot spot' to refer to the region where the mass stream interacts with material already encircling the secondary star and assume that any material which has completed a revolution around the secondary star is part of the disk. 

Even if a true ``hot spot" is not created by the mass stream-disk interaction, the region where the stream and disk meet could potentially decrease observed polarization from the disk edge by disrupting the otherwise smooth structure of the disk edge and adding unpolarized light at phases when it is visible. On the other hand, the mass stream, which is elongated in the same direction as the disk, should produce a polarization position angle very similar to that of the disk. Therefore, the presence of the mass stream should not lead to a decrease in the observed polarization. The effects of a hot spot would be detectable in the polarization light curves in the \textit{BVRI} bands because the disk is the primary scattering region for visible light in the $\beta$ Lyr system. But, if it is not significantly brighter than the surrounding disk, the hot spot would not be visible in the total light curves. Therefore, we interpret the $\%\textit{Q}_p$ minimum associated with the randomization of the polarized position angle prior to secondary eclipse as the first direct evidence for the proposed hot spot on the $\beta$ Lyr disk edge \citep{Lubow,Harmanec2002}.  

We expect the hot spot to create an unstructured region on the disk edge where the polarization vectors of the scattered light are randomized in position angle. When this part of the disk is visible, the hot spot should cause a decreased polarization signal and a rotation in position angle, both of which occur in our polarization curves (Figures \ref{Bband}-\ref{Iband}). As long as the hot spot does not lie on the line connecting the centers of mass of the two stars, and its brightness in the visible continuum is similar to that of the undisturbed disk, its effect should result in a minimum in polarization that does not correspond to a minimum in flux. Hydrodynamical modeling of the $\beta$ Lyr system indicates that the mass stream, and therefore its associated hot spot, should lead the loser in the sense of rotation of the system \citep{Lubow,model}. In the $\beta$ Lyr polarization curves, the polarization minimum and concurrent position angle variation occur just before secondary eclipse, suggesting that the hot spot begins its transit of the disk before the loser does. In fact, the larger dispersion of points in the \textit{B} and \textit{V} bands at the first quadrature phase when compared to the second quadrature phase suggests that the hot spot is already in view by phase 0.25. In this picture, the minimum polarization occurs at the phases where the disk area disrupted by the hot spot and eclipsed by the loser is maximized. We sketch this proposed interpretation in Figure \ref{HotSpotEffect}.

We note that in all bands, there are fewer data points after secondary eclipse than before, which may skew the $\% \textit{Q}_P$ Fourier fit near secondary minimum. However, we have several reasons to believe the eclipse offset is not an artifact of the fit. The effect is apparent in all filters, some of which have a much lower point density in phase then the \textit{V} band. (However, the \textit{V} band displays the smallest difference between the $\% \textit{Q}_P$ near secondary minimum and phase 0.5.) The position angle rotation does not heavily depend on the number of points, is apparent in all bands, and has a larger effect at pre-secondary eclipse phases than post-secondary eclipse phases. Finally, the uncertainties on the phases at which the minima occur are small (see Table \ref{HSSize}) compared to the difference between phase 0.5 and the polarization secondary minimum. Future work will include filling in the post-secondary data gap with new HPOL observations to improve the Fourier fits and quantify the $\% \textit{Q}_P$ near secondary minimum offsets more reliably.

In the subsections below, we outline three different estimates of the size of the hot spot, assuming it has the same height as the edge of the disk. We use the following values for system parameters: a loser radius of $R_L=15R_\sun$, a disk diameter of $D_D=60R_\sun$, a binary separation of $R_S=58R_\sun$, and a disk height of $H_D=16R_\sun$ \citep{Linnell2000, Harmanec2002}.

\subsection{Hot Spot Size Estimate: $\% \textit{Q}_P$ Method}
We can use the offset in secondary eclipse to estimate a maximum size for the hot spot. Assuming circular orbits, we have the scenario depicted in Figure \ref{HotSpotSize}. Knowing that phase 0.5 occurs at an angle of 180$\degree$ on the circle depicting the loser's orbit, we can use a simple ratio to find the angle $\theta$,
\begin{equation}
\frac{0.5}{180\degree}=\frac{P}{180\degree-\theta}
\end{equation}
where $P$ is the phase for which secondary eclipse occurs in polarized light (Table \ref{HSSize}) and $180\degree-\theta$ is the angle from zero at which phase $P$ occurs. If we know $\theta$, we can also find the length of line $x$,
\begin{equation}
x=R_S \sin(\theta)
\end{equation}
where $R_S$ is the radius between the center of the disk and the center of the loser. With the length of line $x$ we can estimate the projected size of the hot spot, $HS_{Q}$ (hatched region in the Observer's View in Figure \ref{HotSpotSize}), with the following equation,
\begin{equation}
HS_{\textit{Q}} = \left\{ \begin{array}{l l}
  60 R_\sun & \mbox{for $x \ge \frac{1}{2}D_D$},\\
  (x-R_L) + \frac{1}{2}D_D & \mbox{for $\frac{1}{2}D_D > x > R_L$}, \\
  \frac{1}{2} D_D - (R_L-x) & \mbox{for $R_L > x > 0$}
\end{array} \right.
\end{equation}
where $D_D$ is the diameter of the accretion disk and $R_L$ is the radius of the loser.

Using the above formulae we can calculate the maximum projected hot spot size for the \textit{BVRI} bands. Table \ref{HSSize} lists the results. The maximum hot spot size ranges from 22 $R_\sun$ to 33 $R_\sun$. Since we assume the hot spot has the same height as the disk, these values represent `widths' along the projected face of the disk. We do not calculate formal error bars on these estimates because the estimates vary so widely. 

\subsection{Hot Spot Size Estimate: Position Angle Method}
We also used the variations in position angle to estimate a maximum size for the hot spot. First, we calculated the size of the disk in phase. To do this we solved Equation 2 for $\theta$ when $x= \frac{1}{2}D_D$. The phase for the left side of the disk as depicted in Figure \ref{HotSpotSize} is then given by Equation 1. We calculate this phase to be 0.413. Similarly for the right side of the disk we calculate a phase of 0.587. The resulting size of the disk in phase is the difference of these phases, or 0.174. 

We then estimated the size of the hot spot in phase by finding how long the randomization of position angle lasts. We assumed any points near secondary eclipse that deviated significantly from the average position angle were due to the hot spot. We did not use a formal calculation to find these points; we chose the smallest and largest phases for the randomized position angles by eye. We then took the difference in phase between the deviant observations with the smallest and largest phases to calculate the size of the hot spot in phase, $\theta_{HS}$ (Table \ref{HSSize}). In this case the maximum hot spot size, $HS_{PA}$, is given by the ratio
\begin{equation}
\frac{60R_\sun}{0.174}=\frac{HS_{PA}}{\theta_{HS}}.
\end{equation} 
Table \ref{HSSize} lists the maximum projected size of the hot spot across the edge of the disk for each band. Our hot spot size estimates found with this method range from 26 $R_\sun$ to 58 $R_\sun$. 

\subsection{Hot Spot Size Estimate: Simple Model}
We used a simple model for a third estimate of the size of the hot spot. For this model we assume the polarization of the disk is uniform across the disk edge. We first calculated a baseline $qf_{DC}$, the polarized flux due to the disk's self-illumination, by taking the error-weighted mean $\%\textit{Q}_P$ multiplied by the normalized Fourier fit flux of the observations between phases 0.7 and 1.2 for each band (HWN). This assumes all the polarized flux at these phases is due to light originating within the disk rather than from the loser, a reasonable assumption given the results of HWN. We also define $qf_{min}$, the minimum polarized flux near secondary eclipse due to the primary star's eclipse and hot spot's transit of the disk, to be the error-weighted mean $\%\textit{Q}_P$ multiplied by the normalized Fourier fit flux for observations between phases 0.4 and 0.55. If we subtract from $qf_{DC}$ the amount of polarized flux blocked by the primary star and disrupted by the transit of the hot spot, the result should be $qf_{min}$, the polarized flux observed at secondary eclipse.

The amount of polarized flux lost due to the primary eclipsing the disk and the hot spot transiting the disk is given by
\begin{equation}
qf_{DC}-\frac{qf_{DC}}{A_D}A_{ecl}-\frac{qf_{DC}}{A_D}A_{HS}=qf_{min}
\end{equation}  
where $A_{ecl}$ is the area of the disk eclipsed by the primary star (shaded region in Figure \ref{HotSpotSize}), $A_D$ is the observed area of the edge of the disk, and $A_{HS}=H_DHS_{SM}$ is the area of the disk edge disrupted by the hot spot. The fraction $qf_{DC}/A_D=qf_{DC}/D_DH_D$ gives the polarized flux per unit area from the disk. The second term is the polarized flux eclipsed by the primary star and the third term is the polarized flux subtracted by the hot spot. We note that Equation 5 assumes an inclination angle of $i=90\degree$ and that each unit area of the disk contributes to the polarized flux equally. In reality, the relative contributions of each portion of a disk with an inclination angle of $i=90\degree$ are not equal due to limb darkening. Taking limb darkening into account would complicate our estimates; a hot spot near a limb darkened edge of the disk would need to be larger to account for the same amount polarized flux loss that would be lost by a hot spot closer to the center of the disk. Table \ref{HSSize} gives $qf_{DC}$, $qf_{min}$, and the resulting hot spot size estimate for the \textit{BVRI} bands using Equation 5. 

Assuming an inclination angle of $i=86\degree$ \citep{Linnell1998, Linnell2000} instead of $90\degree$ changes the equation we use to calculate $HS_{SM}$. In this case, the area of the disk that we see is larger than in the edge-on case; projection effects allow us to see a small portion of the back side of the disk. The visible portion of the interior of the disk is polarized differently than the disk edge; the polarized flux from the interior should cancel with some of the polarized flux from the disk edge. Also, the area eclipsed by the primary star is larger and the center of the star is no longer aligned with the plane that cuts the disk into equal bottom and top halves. 

In order to calculate an estimate for $i=86\degree$, we make the assumption that the area of the disk we see is rectangular. This makes our calculations easier since the projected height of the disk does not change across the disk edge, but has the effect of making our estimate smaller; if the edges of the projected disk have a height of $H_D$ and the center of the projected disk has a larger height due to the projection, the total area of the disk is smaller than a rectangle whose height is the projected height (see Figure \ref{ieffects}). This assumption allows us to use Equation 5 with a slight adjustment:
\begin{equation}
qf_{DC}-\frac{qf_{DC}}{D_DH_{PD}}A_{ecl}-\frac{qf_{DC}}{D_DH_{PD}}A_{HS}=qf_{min}
\end{equation}
where $H_{PD}$ is the projected disk height. A simple calculation reveals that $H_{PD}=20R_\sun$. The area of the hot spot, $A_{HS}=H_DHS_{SM}$, remains unchanged because the hot spot is only on the front portion of the disk. Therefore, it does not take up the full projected disk height. We note that we do not have to account for the cancellation of polarized light due to the contribution from the interior of the disk. We are using our observations to estimate $qf_{DC}$ and $qf_{min}$, and these numbers should therefore already incorporate this effect, if it is present. However, we still make the assumption that each portion of the disk contributes to the polarized flux equally. Besides the complication due to limb darkening mentioned previously, this enlarges our size estimate because more area is needed to cancel out the same amount of polarized flux. Table \ref{HSSize} gives the resulting hot spot size estimate for the \textit{BVRI} bands when $i=86\degree$ using Equation 6.

In our estimates, the size of the hot spot is smaller when the inclination angle is $86\degree$ compared to $90\degree$ for the following reason. The area of the disk is larger in the $i=86\degree$ method compared to the $i=90\degree$ method by a factor of approximately 1.3. This causes the amount of polarized flux per unit area to decrease by a factor of 0.8. However, the area eclipsed by the primary star increases by more than a factor of two. Therefore, the amount of polarized light lost due to the hot spot is smaller when $i=86\degree$ than when $i=90\degree$.

\subsection{Comparison and Review of Hot Spot Size Estimates}

Comparing all three methods, we find a wide range of sizes for the hot spot. The smallest size estimate is 2 R$_\sun$ (\textit{I} band) while the largest is 58 R$_\sun$ (\textit{V} band). The \textit{R} and \textit{I} band estimates are the most likely to change with additional data because their current phase coverage is not as good as the \textit{B} and \textit{V} bands. The \textit{B} band estimates have the closest agreement between the three methods; they range from 22 R$_\sun$ to 33 R$_\sun$, while the \textit{V} band estimates have the largest range, 9 R$_\sun$ to 58 R$_\sun$. The overlap region for size ranges in all the bands is 22 R$_\sun$ to 33 R$_\sun$.

The large size of our estimates lends support to the possibility that we are actually detecting the portion of the mass stream which has not interacted with the disk and not a hot spot, similar to the findings of Bisikalo et al. (2000). The largest hot spot size estimate, 58 R$_\sun$, is similar in size to the diameter of the disk, 60 R$_\sun$, and the same as the binary separation, 58 R$_\sun$. However, this scenario would not likely produce the phenomena seen in Figures \ref{Bband} through \ref{Iband} because light scattering from the mass stream would tend to have a position angle similar to that of the disk; therefore, we prefer a large hot spot interpretation. 

The 22 R$_\sun$ to 33 R$_\sun$ range is likely an upper limit for the size of the hot spot for several reasons. The position angle method (see Section 4.2) relies on using the randomization of position angles around secondary eclipse. We determined the length of time that the randomization in position angle lasts by using data from multiple orbits of the system. If the spot varies, either in size or location, on time scale similar to the orbital period, this variation would cause our estimate to be larger than the actual size of the hot spot. Also, we made the assumption that the hot spot contributes only unpolarized light to the observations. It is possible that the hot spot contributes light polarized at a different position angle than light polarized in the disk. This would cause a cancellation effect to occur; light polarized in the hot spot would cancel some of the light polarized in the disk. In this case, our estimate would again be larger than the area of the hot spot. 

If the hot spot is indeed larger than 30 R$_\sun$, a portion of it may already be visible to the observer during primary eclipse. If this were the case, we would expect the polarization maxima in the affected bands to shift from phase 0.0 to an earlier phase. This line of reasoning suggests that the primary eclipse effects seen in Figures \ref{Bband} through \ref{Iband}, if real, may also be due to the hot spot (see Section 3.1). However, such a scenario does not explain why the \textit{I} band polarization maximum occurs just after primary eclipse. The \textit{R} band and three-term \textit{V} band Fourier fits do not show an offset at primary eclipse, suggesting that the hot spot is not visible at this phase (see Section 3.1). We propose that the primary eclipse offsets are likely a result of the Fourier fits in some bands being less well determined around primary eclipse due to incomplete phase coverage. 

The position angle scatter around primary eclipse in Figures \ref{Bband} through \ref{Iband} could also be interpreted as evidence that the hot spot is visible at this phase. As the system moves from primary eclipse to secondary eclipse, the amount of the hot spot visible to the observer should only increase if the time scale for changes in the hot spot is large compared to the orbital period. This suggests that if the scatter in the position angle of the 1997 August 25 observations (squares between phases 0.000 and 0.015 in Figures \ref{Bband} through \ref{Iband}) is due to the hot spot, then a similar scatter should also exist in observations where the hot spot was fully visible during the same orbit. In particular, the 1997 August 26 (phase 0.079) and 1997 August 30 (phase 0.385) observations should show this effect. However, these two observations appear to have a position angle more similar to the system average than the 1997 August 25 observations. If that the mass transfer rate varies on time scales shorter than an orbital period, then the hot spot size may have changed over this five-day period to change the amount of position angle scatter. This scenario would explain the position angle scatter at primary eclipse for a single cycle, but it would not address the fact that no other orbits of the system were observed to have a position angle scatter near primary eclipse. The combination of the arguments for and against the visibility of the hot spot at primary eclipse does not clearly determine whether its effects are discernible near phase 0.0. Multiple spectropolarimetric observations from the same orbit near primary eclipse will provide more insight into the cause of this phenomena and access its repeatability. In particular, the evolution of the color index of the polarized flux through primary eclipse compared to the evolution of the color index of the total light flux may shed new light on this situation.

Harmanec et al (1996) derived the location of the `roots' of the bipolar outflows, where the bipolar outflows originate within the disk, using the H$\alpha$ absorption core and the H$\alpha$ emission wings seen in many epochs of spectra (see Figure 1 in \cite{Harmanec2002} for an artist's view of the location of the outflows). The location of the roots (marked by a filled square in Figure \ref{HotSpotSize} for the H$\alpha$ absorption and by a filled star for H$\alpha$ emission) is in the same quadrant of the disk in which we interpret the hot spot being located. However, the location of the roots is in the interior of the disk while we suggest the hot spot is a disruption in the structure of the disk edge. Certainly the bipolar outflows and the hot spot are related; both components are the result of the system's high mass transfer rate. How far into the interior of the disk the hot spot reaches is unknown. Additionally, the disk is made up of two components: a dense inner disk and an outer less dense disk \citep{Skulskii}. What constitutes the `disk edge' where the hot spot disruption occurs is unclear, although we have assumed it is the outermost edge of the disk (whose diameter is 60 R$_\sun$) in our size estimates. Because the scattering region for the H$\alpha$ line is thought to be the bipolar outflows (HNF), future high precision H$\alpha$ line polarization measurements may be able to link the two structures. If the hot spot location is consistent with the roots of the outflows, we expect the H$\alpha$ line polarization to show a secondary minimum offset, similar to those characterizing the broadband curves (Figures \ref{Bband} through \ref{Iband}), while maintaining a position angle consistent with the outflows. Radial velocity curves of the H$\alpha$ line's polarized flux may also provide valuable insight into the relationship between the hot spot and bipolar outflows.

\section{SUMMARY}

We have presented a large new data set of polarimetric observations of $\beta$ Lyr in the \textit{BVRI} bands and the first Fourier fits to the polarimetric variations in these bands and the \ion{He}{1} $\lambda$5876 emission line. We have interpreted the minimum in the \textit{BVRI} projected polarization prior to secondary eclipse and the associated position angle rotations as the first direct evidence for a hot spot on the edge of the accretion disk in the $\beta$ Lyr system. Using the phases of polarization minimum, the scatter of the position angle and a simple model, we have estimated the maximum size of the hot spot to be between 22 and 33 R$_\sun$ across the face of the disk. More extensive polarimetric modeling of $\beta$ Lyr is needed in order to fully understand these results. Insights into the importance of the effects at primary eclipse and more accurate estimates of the hot spot size could be derived from such models.

We expect the hot spot may also be detectable in X-rays. Both ROSAT HRI (Berghofer \& Schmitt 1994) and Suzaku (Ignace et al. 2008) have detected strong and variable hard X-ray emission from $\beta$ Lyr. However, neither set of observations has provided information on the origin of the X-ray emission or observed the system at phases at which we see the hot spot effects. An X-ray light curve with more complete phase coverage will help locate the source of the X-ray emission.  

\begin{figure}
\figurenum{6}
\epsscale{1.0}
\plotone{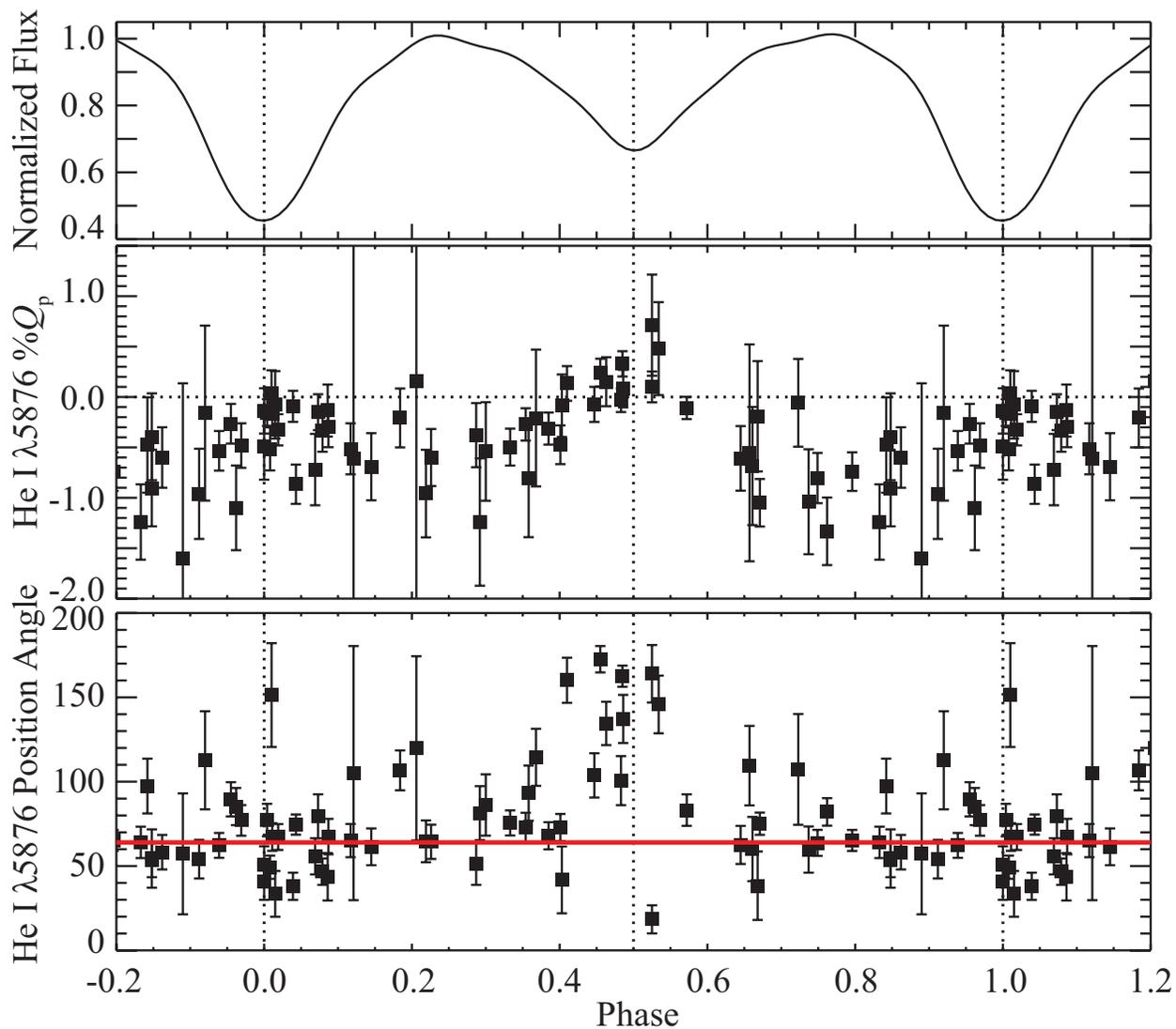}
\label{HeI5876}
\caption{\ion{He}{1} $\lambda$5876 emission line polarization from HPOL CCD observations (see Section 3.2). \textit{From top:} Normalized \textit{V} band Fourier fit light curve \citep{Harmanec1996}, projected polarization, and position angle (degrees) versus phase. Error bars represent intrinsic uncertainties. All data have been wrapped in phase to display more than one complete period.}
\end{figure}

\begin{figure}
\figurenum{7}
\epsscale{1.0}
\plotone{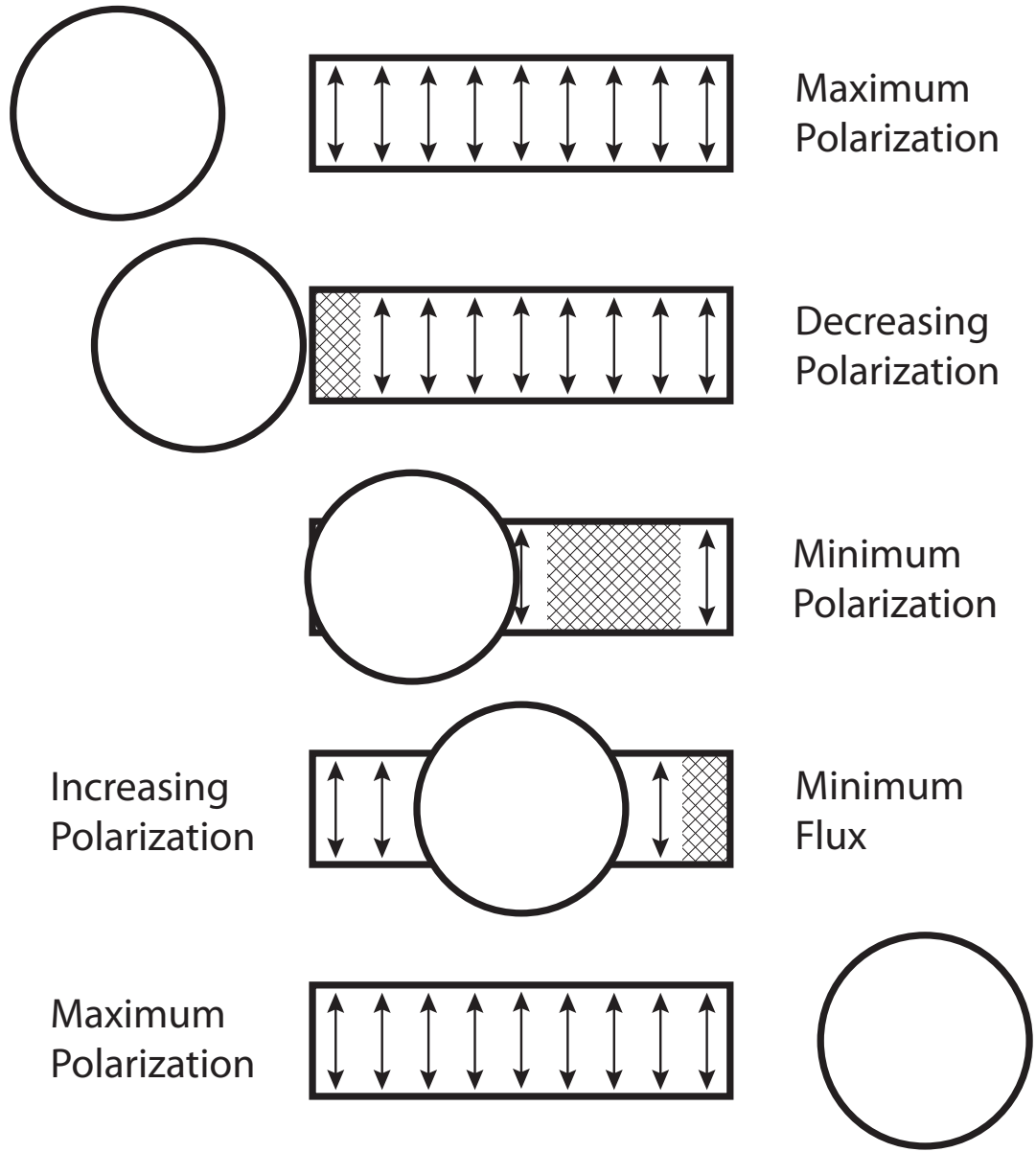}
\label{HotSpotEffect}
\caption{Proposed geometry of the $\beta$ Lyr system at various phases in our proposed hot spot model. Arrows represent the polarization arising from the disk edge. \textit{From Top:} The first maximum in polarization occurs at the first quadrature phase. The polarization then begins to decrease as the hot spot (hatched region) rotates into view. The minimum in polarization occurs when the area eclipsed by the loser and disrupted by the hot spot is maximized. The minimum in flux occurs as the hot spot is rotating off the edge of the visible disk. The second maximum in polarization occurs at the second quadrature phase.}
\end{figure}

\begin{figure}
\figurenum{8}
\epsscale{0.5}
\plotone{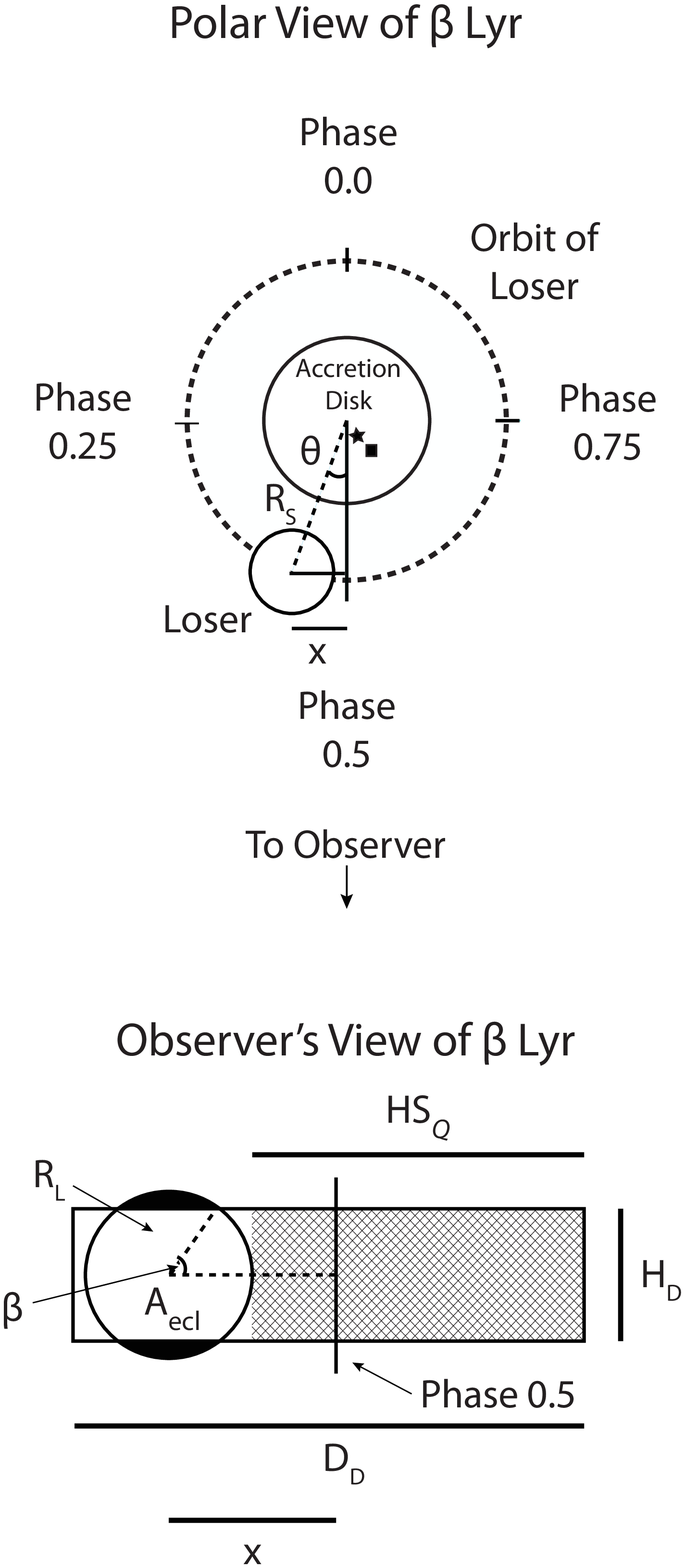}
\label{HotSpotSize}
\caption{Sketch of the geometry we used for our hot spot size estimates (not to scale, see Section 4). We used parameters obtained by Linnell (2000) for the disk height, disk diameter, loser radius and separation between the two components. The hatched region in the Observer's View represents the hot spot size based on the $\%\textit{Q}_p$ method (see Section 4.1). The blackened area of the circle in the Observer's View is the uneclipsed area of the loser at primary eclipse. The filled square inside the accretion disk in the Polar View indicates the location of the roots of the bipolar outflows as given by the H$\alpha$ absorption core while the filled star represents the same thing for the H$\alpha$ emission wings \citep{Harmanec1996}.}
\end{figure}

\begin{figure}
\figurenum{9}
\epsscale{1.0}
\plotone{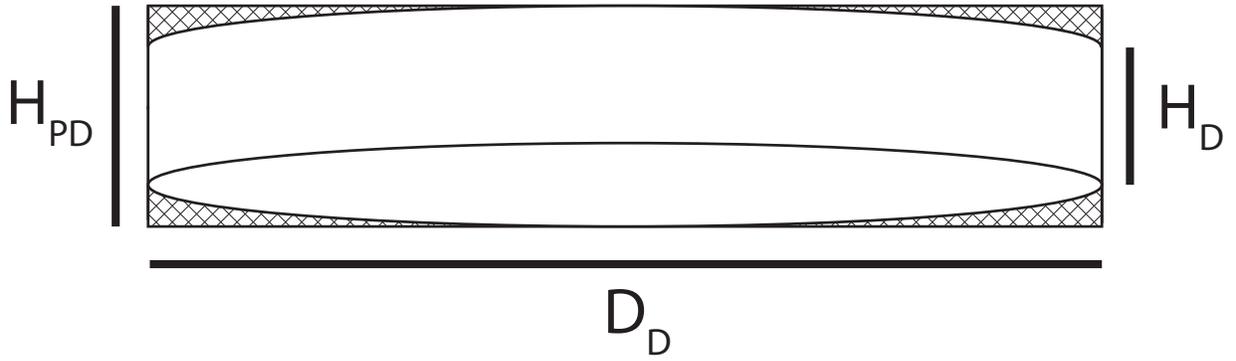}
\label{ieffects}
\caption{Sketch of the geometry we used for our simple model hot spot size estimate at an inclination angle of $86\degree$ (not to scale; see Section 4.3). The hatched regions represent the additional area of the disk in our estimate due to the assumption of a rectangularly shaped disk.}
\end{figure}

The large uncertainties and scatter in Figure \ref{HeI5876} make it difficult to pinpoint a location for the origin of the jets with confidence. Future, higher-precision line polarization measurements will provide much needed insights and determine their source.

Advancements in technology will soon allow for the combination of long-baseline 
optical interferometry with polarimetry \citep{Elias2008}. We expect such a technological development will provide new and exciting geometrical insights into the $\beta$ Lyr system and others like it. 

\acknowledgments
We are very grateful to Brian Babler, Marilyn Meade, and Ken Nordsieck for their help with the HPOL data and to all of the members of the PBO science team for obtaining the HPOL data used in this paper. We are indebted to the late R.H. Koch for his observing efforts at FCO and to his wife Joanne for allowing us to use those data. We also thank Keivan Stassun for his help with the period analysis, and Petr Harmanec for providing us with unpolarized \textit{UBVRI} light curves of $\beta$ Lyr and his helpful referee comments. We have had useful conversations with many colleagues: Kathleen Geise, Paul Hemenway, K. Tabetha Hole, Richard Ignace, Brian Kloppenborg, Robert Stencel, and Toshyia Ueta. JRL acknowledges support through the NASA Harriett G. Jenkins Pre-doctoral Fellowship Program and Sigma Xi's Grants-in-Aid of Research Program. JLH acknowledges support though NASA ADP award NNH08CD10C and NSF award AST-0807477. FAB acknowledges support from NSF grant AST-0849736 (K. Stassun, PI) and from a GAANN Fellowship.

\clearpage

\begin{deluxetable}{lcc}
\tablewidth{0pt}
\tablecaption{Date and Phase Information for Midpoints of the HPOL $\beta$ Lyrae and $\beta$ Lyrae B Observations \label{betLyrObs}}
\tablehead{
\colhead{Date}           &\colhead{HJD - 2,400,000}      &
\colhead{Phase\tablenotemark{a}} }
\startdata
\multicolumn{3}{l}{$\beta$ Lyr Reticon (through HNF; recalibrated):}\\
\hspace{3mm} 1992 Sep 30		&48895.71	&0.661			\\
\hspace{3mm} 1992 Oct 6			&48901.67	&0.121			\\
\hspace{3mm} 1992 Oct 13		&48908.60	&0.657			\\
\hspace{3mm} 1992 Oct 27		&48922.58	&0.737			\\
\hspace{3mm} 1992 Dec 28		&48984.52	&0.525			\\
\hspace{3mm} 1993 Jul 26		&49194.60	&0.762			\\
\hspace{3mm} 1994 Jun 3			&49506.78	&0.890			\\
\enddata
\tablenotetext{a}{Phases were calculated using the ephemeris in Harmanec \& Scholz (1993).}
\tablenotetext{b}{Table 1 of HNF incorrectly lists this date as 1994 March 14.}
\tablenotetext{c}{These observations used only the red grating; see Section 2.1.}
\tablenotetext{d}{These observations used only the blue grating; see Section 2.1.}
\end{deluxetable}

\begin{deluxetable}{lcc}
\tablewidth{0pt}
\tablecaption{Date and Phase Information for Midpoints of the FCO $\beta$ Lyrae Observations \label{betLyrObsFCO}}
\tablehead{
\colhead{Date}           &\colhead{HJD - 2,400,000}      &
\colhead{Phase\tablenotemark{a}} }
\startdata
\multicolumn{3}{l}{\textit{B} Band:}\\
    1989 Apr 12    & 47628.88 & 0.753 \\
    1989 Apr 20    & 47636.85 & 0.352 \\
    1989 Apr 23    & 47639.85 & 0.583 \\
    1989 May 29    & 47675.80 & 0.366 \\
    1989 Jun 5     & 47682.78 & 0.907 \\
    1989 Jun 12    & 47689.80 & 0.448 \\
    1989 Jun 19    & 47696.78 & 0.989 \\
\enddata
\tablenotetext{a}{Phases were calculated using the ephemeris in Harmanec \& Scholz (1993).}
\end{deluxetable}

\begin{deluxetable}{lccccccc}
\tablecolumns{10}
\tablewidth{0pt}
\tabletypesize{\scriptsize}
\tablecaption{FCO \textit{B} Filter Data \label{BFCOData}}
\tablehead{
\colhead{Phase} & \colhead{HJD - 2,400,000} & \colhead{\%\textit{Q}} & \colhead{\%\textit{U}} & \colhead{ISP sub $\% \textit{Q}$} & \colhead{{ISP sub $\% \textit{U}$ }} & \colhead{$\sigma_{i,Q}$\tablenotemark{a} (\%)} & \colhead{$\sigma_{i,U}$\tablenotemark{a} (\%)}}
\startdata
    0.081 & 47736.79 & 0.395 & -0.432 & 0.164 & -0.067 & 0.007 & 0.007 \\
    0.308 & 47778.57 & 0.579 & -0.493 & 0.348 & -0.128 & 0.016 & 0.014 \\
    0.318 & 47804.45 & 0.617 & -0.527 & 0.386 & -0.162 & 0.010 & 0.008 \\
    0.352 & 47636.85 & 0.405 & -0.438 & 0.174 & -0.073 & 0.008 & 0.008 \\
    0.366 & 47675.80 & 0.549 & -0.520 & 0.318 & -0.155 & 0.009 & 0.008 \\
    0.448 & 47689.80 & 0.257 & -0.261 & 0.026 & 0.104 & 0.007 & 0.007 \\
    0.543 & 48777.76 & 0.148 & -0.350 & -0.083 & 0.015 & 0.009 & 0.009 \\
    0.550 & 47807.60 & 0.460 & -0.698 & 0.229 & -0.333 & 0.016 & 0.023 \\
    0.583 & 47639.85 & 0.421 & -0.527 & 0.190 & -0.162 & 0.008 & 0.010 \\
    0.632 & 47769.68 & 0.497 & -0.517 & 0.266 & -0.152 & 0.010 & 0.010 \\
    0.753 & 47628.88 & 0.538 & -0.479 & 0.307 & -0.114 & 0.014 & 0.012 \\
    0.772 & 47732.74 & 0.486 & -0.494 & 0.255 & -0.129 & 0.011 & 0.011 \\
    0.840 & 47707.76 & 0.416 & -0.469 & 0.185 & -0.104 & 0.007 & 0.008 \\
    0.844 & 47772.58 & 0.406 & -0.476 & 0.175 & -0.111 & 0.009 & 0.011 \\
    0.907 & 47682.78 & 0.531 & -0.488 & 0.300 & -0.123 & 0.012 & 0.011 \\
    0.922 & 47773.57 & 0.470 & -0.456 & 0.239 & -0.091 & 0.010 & 0.010 \\
    0.989 & 47696.78 & 0.602 & -0.566 & 0.371 & -0.201 & 0.010 & 0.010 \\
    0.994 & 47709.79 & 0.635 & -0.558 & 0.404 & -0.193 & 0.018 & 0.016 \\
    0.999 & 47774.57 & 0.553 & -0.645 & 0.322 & -0.280 & 0.009 & 0.010 \\
\enddata
\tablenotetext{a}{The $\sigma_{i,Q}$ and $\sigma_{i,U}$ columns represent internal errors.}
\end{deluxetable}

\begin{deluxetable}{lccccccc}
\tablecolumns{10}
\tablewidth{0pt}
\tabletypesize{\scriptsize}
\tablecaption{FCO \textit{V} Filter Data \label{VFCOData}}
\tablehead{
\colhead{Phase} & \colhead{HJD - 2,400,000} & \colhead{\%\textit{Q}} & \colhead{\%\textit{U}} & \colhead{ISP sub $\% \textit{Q}$} & \colhead{{ISP sub $\% \textit{U}$ }} & \colhead{$\sigma_{i,Q}$\tablenotemark{a} (\%)} & \colhead{$\sigma_{i,U}$\tablenotemark{a} (\%)}}
\startdata
  0.002 & 48822.60 & 0.353 & -0.393 & 0.134 & -0.028 & 0.053 & 0.053 \\
    0.017 & 48809.68 & 0.405 & -0.471 & 0.186 & -0.106 & 0.013 & 0.013 \\
    0.018 & 47412.55 & 0.508 & -0.433 & 0.289 & -0.068 & 0.010 & 0.009 \\
    0.037 & 47464.50 & 0.687 & -0.704 & 0.468 & -0.339 & 0.097 & 0.099 \\
    0.081 & 47736.82 & 0.390 & -0.422 & 0.171 & -0.057 & 0.010 & 0.011 \\
    0.084 & 48836.61 & 0.390 & -0.423 & 0.171 & -0.058 & 0.010 & 0.010 \\
    0.104 & 47025.54 & 0.287 & -0.412 & 0.068 & -0.047 & 0.014 & 0.019 \\
    0.104 & 47025.56 & 0.292 & -0.437 & 0.073 & -0.072 & 0.008 & 0.012 \\
\enddata
\tablenotetext{a}{The $\sigma_{i,Q}$ and $\sigma_{i,U}$ columns represent internal errors.}
\end{deluxetable}

\begin{deluxetable}{lccccccc}
\tablecolumns{10}
\tablewidth{0pt}
\tabletypesize{\scriptsize}
\tablecaption{FCO \textit{R} Filter Data \label{RFCOData}}
\tablehead{
\colhead{Phase} & \colhead{HJD - 2,400,000} & \colhead{\%\textit{Q}} & \colhead{\%\textit{U}} & \colhead{ISP sub $\% \textit{Q}$} & \colhead{{ISP sub $\% \textit{U}$ }} & \colhead{$\sigma_{i,Q}$\tablenotemark{a} (\%)} & \colhead{$\sigma_{i,U}$\tablenotemark{a} (\%)}}
\startdata
    0.081 & 47736.84 & 0.318 & -0.373 & 0.131 & -0.077 & 0.007 & 0.008 \\
    0.105 & 47801.65 & 0.454 & -0.432 & 0.267 & -0.136 & 0.011 & 0.011 \\
    0.308 & 47778.62 & 0.503 & -0.431 & 0.316 & -0.135 & 0.008 & 0.007 \\
    0.352 & 47636.81 & 0.376 & -0.374 & 0.189 & -0.078 & 0.012 & 0.012 \\
    0.366 & 47675.76 & 0.504 & -0.489 & 0.317 & -0.193 & 0.018 & 0.017 \\
    0.448 & 47689.84 & 0.261 & -0.267 & 0.074 & 0.029 & 0.006 & 0.007 \\
    0.583 & 47639.79 & 0.377 & -0.450 & 0.190 & -0.154 & 0.009 & 0.011 \\
    0.632 & 47769.72 & 0.412 & -0.443 & 0.225 & -0.147 & 0.013 & 0.014 \\
    0.704 & 47809.61 & 0.830 & -0.779 & 0.643 & -0.483 & 0.017 & 0.016 \\
    0.753 & 47628.90 & 0.452 & -0.417 & 0.265 & -0.121 & 0.014 & 0.013 \\
    0.772 & 47732.78 & 0.398 & -0.398 & 0.211 & -0.102 & 0.009 & 0.009 \\
    0.840 & 47707.80 & 0.359 & -0.373 & 0.172 & -0.077 & 0.009 & 0.004 \\
    0.844 & 47772.62 & 0.330 & -0.372 & 0.143 & -0.076 & 0.007 & 0.007 \\
    0.907 & 47682.83 & 0.391 & -0.443 & 0.204 & -0.147 & 0.012 & 0.021 \\
    0.989 & 47696.82 & 0.449 & -0.417 & 0.262 & -0.121 & 0.011 & 0.010 \\
    0.994 & 47709.83 & 0.524 & -0.438 & 0.337 & -0.142 & 0.017 & 0.015 \\
    0.999 & 47774.62 & 0.463 & -0.522 & 0.276 & -0.226 & 0.013 & 0.014 \\
\enddata
\tablenotetext{a}{The $\sigma_{i,Q}$ and $\sigma_{i,Q}$ columns represent internal errors.}
\end{deluxetable}

\begin{deluxetable}{lccccccc}
\tablecolumns{10}
\tablewidth{0pt}
\tabletypesize{\scriptsize}
\tablecaption{HPOL Synthetic \textit{B} Filter Data \label{BHPOLData}}
\tablehead{
\colhead{Phase} & \colhead{HJD - 2,400,000} & \colhead{\%\textit{Q}} & \colhead{\%\textit{U}} & \colhead{ISP sub $\% \textit{Q}$} & \colhead{{ISP sub $\% \textit{U}$ }} & \colhead{$\sigma_i$\tablenotemark{a} (\%)} & \colhead{$\sigma_s$\tablenotemark{a} (\%)}}
\startdata
\multicolumn{3}{l}{\textit{B} Band Reticon:} \\
    0.000 & 49663.46 & 0.4323 & -0.6592 & 0.2020 & -0.2953 & 0.0019 & 0.0200 \\
    0.121 & 48901.67 & 0.3905 & -0.4928 & 0.1603 & -0.1289 & 0.0030 & 0.0200 \\
    0.206 & 49562.62 & 0.4691 & -0.4456 & 0.2388 & -0.0816 & 0.0062 & 0.0200 \\
    0.292 & 49615.49 & 0.4248 & -0.5536 & 0.1945 & -0.1897 & 0.0015 & 0.0200 \\
    0.358 & 49564.59 & 0.4874 & -0.4409 & 0.2572 & -0.0770 & 0.0023 & 0.0200 \\
    0.368 & 49603.53 & 0.4823 & -0.4503 & 0.2521 & -0.0864 & 0.0016 & 0.0200 \\
    0.525 & 48984.52 & 0.2897 & -0.4828 & 0.0594 & -0.1188 & 0.0034 & 0.0200 \\
\enddata
\tablenotetext{a}{The $\sigma_{i}$ and $\sigma_{s}$ columns represent internal and systematic errors respectively.}
\end{deluxetable}

\begin{deluxetable}{lccccccc}
\tablecolumns{10}
\tablewidth{0pt}
\tabletypesize{\scriptsize}
\tablecaption{HPOL Synthetic \textit{V} Filter Data \label{VHPOLData}}
\tablehead{
\colhead{Phase} & \colhead{HJD - 2,400,000} & \colhead{\%\textit{Q}} & \colhead{\%\textit{U}} & \colhead{ISP sub $\% \textit{Q}$} & \colhead{{ISP sub $\% \textit{U}$ }} & \colhead{$\sigma_i$\tablenotemark{a} (\%)} & \colhead{$\sigma_s$\tablenotemark{a} (\%)}}
\startdata
\multicolumn{3}{l}{\textit{V} Band Reticon:} \\
    0.000 & 49663.46 & 0.4485 & -0.6204 & 0.2292 & -0.2737 & 0.0021 & 0.0200 \\
    0.121 & 48901.67 & 0.3620 & -0.4839 & 0.1427 & -0.1373 & 0.0025 & 0.0200 \\
    0.206 & 49562.62 & 0.4578 & -0.3863 & 0.2386 & -0.0398 & 0.0056 & 0.0200 \\
    0.292 & 49615.49 & 0.4058 & -0.5298 & 0.1865 & -0.1830 & 0.0018 & 0.0200 \\
    0.358 & 49564.59 & 0.4499 & -0.4464 & 0.2306 & -0.0998 & 0.0024 & 0.0200 \\
    0.368 & 49603.53 & 0.4474 & -0.4376 & 0.2280 & -0.0909 & 0.0020 & 0.0200 \\
    0.525 & 48984.52 & 0.2534 & -0.4453 & 0.0339 & -0.0985 & 0.0026 & 0.0200 \\
\enddata
\tablenotetext{a}{The $\sigma_{i}$ and $\sigma_{s}$ columns represent internal and systematic errors respectively.}
\end{deluxetable}

\begin{deluxetable}{lccccccc}
\tablecolumns{10}
\tablewidth{0pt}
\tabletypesize{\scriptsize}
\tablecaption{HPOL Synthetic \textit{R} Filter Data \label{RHPOLData}}
\tablehead{
\colhead{Phase} & \colhead{HJD - 2,400,000} & \colhead{\%\textit{Q}} & \colhead{\%\textit{U}} & \colhead{ISP sub $\% \textit{Q}$} & \colhead{{ISP sub $\% \textit{U}$ }} & \colhead{$\sigma_i$\tablenotemark{a} (\%)} & \colhead{$\sigma_s$\tablenotemark{a} (\%)}}
\startdata
\multicolumn{3}{l}{\textit{R} Band Reticon:} \\
    0.000 & 49663.46 & 0.4242 & -0.5664 & 0.2205 & -0.2422 & 0.0027 & 0.0200 \\
    0.121 & 48901.67 & 0.3360 & -0.4694 & 0.1341 & -0.1494 & 0.0038 & 0.0200 \\
    0.206 & 49562.62 & 0.3777 & -0.3521 & 0.1861 & -0.0332 & 0.0067 & 0.0200 \\
    0.292 & 49615.49 & 0.3735 & -0.4713 & 0.1709 & -0.1498 & 0.0025 & 0.0200 \\
    0.358 & 49564.59 & 0.3955 & -0.4077 & 0.1947 & -0.0863 & 0.0032 & 0.0200 \\
    0.368 & 49603.53 & 0.3917 & -0.4168 & 0.1852 & -0.0923 & 0.0027 & 0.0200 \\
    0.525 & 48984.52 & 0.2297 & -0.4119 & 0.0295 & -0.0905 & 0.0032 & 0.0200 \\
\enddata
\tablenotetext{a}{The $\sigma_{i}$ and $\sigma_{s}$ columns represent internal and systematic errors respectively.}
\end{deluxetable}

\begin{deluxetable}{lccccccc}
\tablecolumns{10}
\tablewidth{0pt}
\tabletypesize{\scriptsize}
\tablecaption{HPOL Synthetic \textit{I} Filter Data \label{IHPOLData}}
\tablehead{
\colhead{Phase} & \colhead{HJD - 2,400,000}  & \colhead{\%\textit{Q}} & \colhead{\%\textit{U}} & \colhead{ISP sub $\% \textit{Q}$} & \colhead{{ISP sub $\% \textit{U}$ }} & \colhead{$\sigma_i$\tablenotemark{a} (\%)} & \colhead{$\sigma_s$\tablenotemark{a} (\%)}}
\startdata
\multicolumn{3}{l}{\textit{I} Band Reticon:} \\
    0.000 & 49663.46 & 0.3626 & -0.5178 & 0.1859 & -0.1806 & 0.0107 & 0.0200 \\
    0.121 & 48901.67 & 0.3121 & -0.5255 & 0.1643 & -0.2718 & 0.0210 & 0.0200 \\
    0.206 & 49562.62 & 0.2211 & -0.3197 & 0.2852 & -0.0712 & 0.0212 & 0.0200 \\
    0.292 & 49615.49 & 0.2800 & -0.4636 & 0.1333 & -0.1712 & 0.0147 & 0.0200 \\
    0.358 & 49564.59 & 0.2809 & -0.3208 & 0.1720 & -0.0486 & 0.0135 & 0.0200 \\
    0.368 & 49603.53 & 0.3388 & -0.4346 & 0.1126 & -0.0753 & 0.0167 & 0.0200 \\
    0.525 & 48984.52 & 0.2337 & -0.3604 & 0.1426 & -0.1002 & 0.0127 & 0.0200 \\
\enddata
\tablenotetext{a}{The $\sigma_{i}$ and $\sigma_{s}$ columns represent internal and systematic errors respectively.}
\end{deluxetable}

\begin{deluxetable}{lcccc}
\tablewidth{0pt}
\tablecaption{Position Angle Information for $\beta$ Lyrae \label{betLyrPA}}
\tablehead{
\colhead{Band}           & \colhead{HPOL Reticon (deg)}      &
\colhead{HPOL CCD (deg)} & \colhead{AH (deg)} & \colhead{FCO (deg)} }
\startdata
\textit{B}	&139.8 $\pm$ 3.5	&165.0 $\pm$ 0.7	&169.8 $\pm$ 1.7 &165.1 $\pm$ 0.9\\
\textit{V}	&147.3 $\pm$ 3.0	&164.0 $\pm$ 0.5	&171.8 $\pm$ 1.0 &163.1 $\pm$ 0.4\\
\textit{R}	&158.4 $\pm$ 0.4	&158.3 $\pm$ 0.4	&\nodata &162.2 $\pm$ 0.7\\
\textit{I}	&57.3 $\pm$ 3.0	&162.7 $\pm$ 0.6	&\nodata &\nodata \\
Index	&	158.8 $\pm$ 0.6	&142.1 $\pm$ 1.7	&\nodata &\nodata \\
H$\alpha$ & \nodata &47.9 $\pm$ 9.7 &\nodata &\nodata\\
H$\beta$ & \nodata &71.1 $\pm$ 2.5 &\nodata &\nodata\\
\ion{He}{1}$\lambda$5876 & \nodata &63.8 $\pm$ 6.4 &\nodata &\nodata\\
\ion{He}{1}$\lambda$7065 & \nodata &51.0 $\pm$ 3.2 &\nodata &\nodata\\
\ion{He}{1}$\lambda$6678 & \nodata &138.2 $\pm$ 6.6 &\nodata &\nodata\\
\enddata
\tablecomments{These values were calculated by fitting a line in \textit{Q-U} space to the data, excluding points between phases 0.475 and 0.525. For the HPOL data, we used systematic errors in both \textit{Q} and \textit{U} in the least-squares fit calculation for all bands except the Balmer jump index (the vector difference between the polarization above and below the Balmer jump) and the lines, for which we used the intrinsic errors. See Sections 3.1 and 3.2.}
\end{deluxetable}

\begin{deluxetable}{lccccc}
\tablewidth{0pt}
\tablecaption{\textit{BVRI} Fourier Fit Parameters \label{FitParam}}
\tablehead{\colhead{Band}  &\colhead{Zero Point} &\colhead{Term} &\colhead{$\Omega$} & \colhead{\textit{A}} &\colhead{$\phi$}}
\startdata
     \textit{B}     & 0.239470 & 1     & 2.98  $\pm$ 0.01  & 0.127 $\pm$ 0.009 & 0.30  $\pm$ 0.01 \\
          & \nodata  & 2     & 1.00  $\pm$ 0.02  & 0.078 $\pm$ 0.008 & 0.37  $\pm$ 0.02 \\
    \textit{V}     & 0.243419 & 1     & 2.98  $\pm$ 0.01  & 0.096 $\pm$ 0.007 & 0.30  $\pm$ 0.01 \\
          & \nodata  & 2     & 1.01  $\pm$ 0.03  & 0.050 $\pm$ 0.006 & 0.36  $\pm$ 0.02 \\
    \textit{R}     & 0.241148 & 1     & 3.00  $\pm$ 0.01  & 0.107 $\pm$ 0.008 & 0.32  $\pm$ 0.01 \\
          & \nodata  & 2     & 4.02  $\pm$ 0.03  & 0.052 $\pm$ 0.009 & 0.08  $\pm$ 0.03 \\
    \textit{I}     & 0.195021 & 1     & 3.00  $\pm$ 0.01  & 0.093 $\pm$ 0.006 & 0.30  $\pm$ 0.01 \\
          & \nodata  & 2     & 4.01  $\pm$ 0.02  & 0.050 $\pm$ 0.006 & 0.10  $\pm$ 0.02 \\
\enddata
\tablecomments{These parameters describe our Fourier fits to the broadband polarimetric data in the combined HPOL, FCO and AH data sets. See Section 3.1.}
\end{deluxetable}

\newpage

\begin{deluxetable}{lccccccccccc}
\rotate
\tablecolumns{11}
\tablewidth{0pt}
\tabletypesize{\scriptsize}
\tablecaption{Broadband Secondary Eclipse Phases and Maximum Hot Spot Size\label{HSSize}}
\tablehead{
\colhead{Band} & \colhead{Phase} & \colhead{Min PA Phase Rot} & \colhead{Max PA Phase Rot} & \colhead{Hot Spot Size in Phase} &\colhead{$qf_{DC}$} & \colhead{$qf_{min}$} & \colhead{$HS_{\textit{Q}}$} & \colhead{$HS_{PA}$} & \colhead{$HS_{SM}$ ($i=90\degree$) } & \colhead{$HS_{SM}$ ($i=86\degree$)}}
\startdata
\textit{B} & 0.481 $\pm$ 0.004  & 0.448 & 0.543 & 0.095 &0.186 &0.045& 22 $R_\sun$& 33 $R_\sun$  &31 $R_\sun$&26 $R_\sun$\\
\textit{V} & 0.481 $\pm$ 0.004 & 0.385 & 0.554 & 0.169 &0.184 &0.080& 22 $R_\sun$ & 58 $R_\sun$ &18 $R_\sun$&9 $R_\sun$\\
\textit{R} & 0.449 $\pm$ 0.005& 0.448 & 0.524 & 0.076  &0.171 &0.080& 33 $R_\sun$& 26 $R_\sun$ &19 $R_\sun$&10 $R_\sun$\\
\textit{I} & 0.449 $\pm$ 0.004& 0.448 & 0.524 & 0.076 &0.140 &0.077& 33 $R_\sun$ & 26 $R_\sun$ &12 $R_\sun$&2 $R_\sun$\\
\enddata
\tablecomments{Columns two through six are parameters used in the three hot spot size estimates (see Sections 4.1, 4.2, and 4.3). Flux normalized to maximum light was used for $qf_{DC}$ and $qf_{min}$. The last four columns represent the width of the hot spot calculated using our different estimates (see Section 4). All three polarimetric data sets were used in this analysis and the resulting estimates.}
\end{deluxetable}

\clearpage

\end{document}